\documentclass[12pt]{article}
\usepackage{graphicx}
 \usepackage{longtable}
\relax
\usepackage{amssymb}

\newcommand{\bo}{{\bar o}}

\def\bo{{\raise.15ex\hbox{\large$\Box$}}}               

\def\face{{\raise.2ex\hbox{$\displaystyle \bigodot$}\mskip-2.2mu \llap {$\ddot
        \smile$}}}                                      

\def\Zbf{{\bf Z}}

\def\leftrightarrowfill{$\mathsurround=0pt \mathord\leftarrow \mkern-6mu
        \cleaders\hbox{$\mkern-2mu \mathord- \mkern-2mu$}\hfill
        \mkern-6mu \mathord\rightarrow$}       
\def\dvec#1{\vbox{\ialign{##\crcr
        \leftrightarrowfill\crcr\noalign{\kern-1pt\nointerlineskip}
        $\hfil\displaystyle{#1}\hfil$\crcr}}}           

\def\beq{\begin{equation}}
\def\eeq{\end{equation}}

\def\beqx{\begin{displaymath}}
\def\eeqx{\end{displaymath}}

\def\beql{\begin{eqnarray}}
\def\eeql{\end{eqnarray}}

\newcommand{\bea}{\begin{eqnarray}}
\newcommand{\eea}{\end{eqnarray}}

\newcommand{\mod}{\;{\rm mod }\;}

\def\[{\left [}
\def\]{\right ]}
\def\({\left (}
\def\){\right )}

\def\+{\oplus}

\begin{document}

\hbox{\hskip 12cm NIKHEF/2010-08  \hfil}
\hbox{\hskip 12cm IFF-FM-2010/01  \hfil}
\hbox{\hskip 12cm March 2010  \hfil}

\vskip .5in

\begin{center}
{\Large \bf Asymmetric Gepner Models\\\vskip .5 truecm
 (Revisited) }

\vspace*{.4in}
{ B. Gato-Rivera}$^{a,b}$
{and A.N. Schellekens}$^{a,b,c}$
\\
\vskip .2in

${ }^a$ {\em NIKHEF Theory Group, Kruislaan 409, \\
1098 SJ Amsterdam, The Netherlands} \\

\vskip .2in

${ }^b$ {\em Instituto de F\'\i sica Fundamental, CSIC, \\
Serrano 123, Madrid 28006, Spain} \\

\vskip .2in

${ }^c$ {\em IMAPP, Radboud Universiteit,  Nijmegen}

\end{center}

\vspace*{0.3in}
{\small
We reconsider a class of heterotic string theories studied in 1989, based on
tensor products of $N=2$ minimal models with asymmetric simple current invariants.
We extend this analysis from $(2,2)$ and $(1,2)$ spectra to 
$(0,2)$ spectra with $SO(10)$ broken to the Standard Model.
In the latter case the spectrum must contain fractionally charged particles. We find that in nearly all cases
at least some of them are massless.
However, we identify a large subclass
where the fractional charges are at worst half-integer, and often vector-like. The
number of families is very often reduced in comparison to the
1989 results, but there are no new tensor combinations yielding three families. All tensor
combinations turn out to fall into two classes: those where the number of families is {\it always} divisible
by three, and those where it is {\it never}  divisible by three. We find an empirical rule to
determine the class, which appears to extend beyond minimal $N=2$ tensor products. 
We observe that distributions of physical quantities such as the number of families, singlets and mirrors have
an interesting tendency towards smaller values as the gauge groups approaches the Standard Model.
We compare our results with an analogous class of free fermionic models. This displays similar
features, but with less resolution.
Finally we present a complete scan of
the three family models based on the triply-exceptional combination $(1,16^*,16^*,16^*)$ identified
originally by Gepner. We find 1220 distinct three family spectra in this case, forming
610 mirror pairs. About half  of them have
the gauge group $SU(3)\times SU(2)_L \times SU(2)_R \times U(1)^5$, the theoretical minimum, and many others 
are trinification models. }


\vskip 1in

\noindent
\newpage

\section{Introduction}

In 1989 one of us co-authored a paper \cite{Schellekens:1989wx} on tensor products of $N=2$ minimal models (also known as ``Gepner models"), applied to the construction of heterotic strings.
In this paper the original construction of Gepner \cite{Gepner:1987qi} was generalized from diagonal (or charge conjugation)
modular invariants   to products of simple current invariants. There have been
other studies of such generalizations \cite{Lutken:1988hc,Fuchs:1989yv}, but they were limited to 
$(2,2)$ conformal field theories
with families in the representation $(27)$ of $E_6$. In \cite{Schellekens:1989wx} some other ideas were considered: the possibility
to break space-time and world-sheet supersymmetry in the bosonic (left-moving)  sector of the heterotic string, and the possibility to
break the gauge group $E_6$ to $SO(10)$, $SU(5)$ or $SU(3)\times SU(2)\times U(1)$. Of all these possibilities, only
one was systematically studied\footnote{The results, referred to as $(1,2)$-models in \cite{Schellekens:1989wx}, are available as scanned pdf files via the  website
www.nikhef.nl/$\sim$t58, on the page ``Hodge numbers".  They were obtained using a nearly
saturated random search.  On the same page one can find a recent
complete enumeration of all (2,2) Gepner models obtained by means of simple currents.}  for all 168 combinations of minimal models, namely breaking left-moving
space-time supersymmetry, which results in breaking $E_6$ to $SO(10)$. The additional breaking of world-sheet supersymmetry was only
studied in one example, namely the $(1,16^*,16^*,16^*)$ Gepner model, where the asterisk indicates the use of an exceptional
modular invariant. This tensor product was singled out because it was the only one known to give rise to spectra with
three families \cite{Gepner:1987hi} (see also \cite{Schimmrigk:1987ke} for the corresponding Calabi-Yau construction). The breaking of $SO(10)$ to the Standard Model was merely mentioned in the conclusions
of \cite{Schellekens:1989wx}, where the authors said about these options: ``We hope to come back to this in the future".

Indeed, the issue of $SO(10)$ breaking to the Standard Model was discussed shortly after in  \cite{Schellekens:1989qb}, and
led to the disappointing conclusion that the spectrum of such theories necessarily contains fractionally charged particles\rlap.\footnote{Throughout this paper, ``fractional charge" refers to electric charges of elementary or
QCD-composite color singlets. In other words, to representations of the Lie algebra $SU(3)\times SU(2) \times U(1)$ that are not representations of the group $S(U(3)\times U(2))$.} This generalized an earlier
result  \cite{Wen:1985qj} for Calabi-Yau compactifications. The appearance of fractionally charged particles  is not necessarily a phenomenological disaster, since such 
fractionally charged particles may be massive, or confined by non-SM interactions, but it does destroy the beautiful understanding of
absence of fractional charges based on $SU(5)$ or $SO(10)$ GUT models.

The purpose of the present paper is to return to the work left unfinished in 1989, and consider these issues again
in the light of new insights, new theoretical methods as well as vastly improved computational power. 

An important change in perspective that has taken place since then is certainly what is now called the string theory landscape \cite{ALS}
\cite{DutchText}.
While
it was clear to some people already a few years before 1989 that string theory was not leading to a unique gauge theory, and
that this was a feature, and not a bug \cite{Linde:1986fd,Schellekens:1987zy}, the full scope of the set of possibilities was
not understood (and may well be largely unknown even now).  Meanwhile several new fertile areas in the landscape
have been discovered and explored, most notably orientifolds (for a review  see \cite{Blumenhagen:2005mu})
and F-theory \cite{Donagi:2008ca,Beasley:2008dc}. 

Although these developments make it more likely that the Standard Model (as we know it today)  can  actually be realized in terms
of string theory or M-theory, perhaps paradoxically, they also cast some doubts upon the original evidence in favor of string theory as a theory of all interactions. In 1984, the encouraging discovery was that Grand Unified Theories emerge naturally from heterotic strings, merely
by compactifying them to four dimensions. However, we know now that heterotic strings are only part of the story, and that in all
other string constructions GUTs only arise because one imposes GUT unification
as a condition. Consider for example orientifold models. In a subclass of these models
(characterized by ``$x=0$" in the terminology of \cite{adks}), $SU(5)$ GUTs are possible. However, instead
of putting five branes on top of each other to get $SU(5)$, one can also take two stacks of three and two at different positions.
Not surprisingly, the latter configuration is
 realized much more frequently.
In another class of orientifold models, the $x=\frac12$ models that include the 
``Madrid" configuration \cite{Ibanez:2001nd} and that dominate the RCFT orientifold landscape, approximate gauge coupling unification can only
arise as a coincidence.

So there is some tension between GUTs and strings.
On the one hand, the aforementioned result on charge quantization \cite{Schellekens:1989qb} ruins 
the nice GUT explanation precisely in the one context were Grand Unification appears to arise naturally in string theory: the heterotic string. 
But in other string constructions, such as $x=0$ orientifolds, F-theory, or higher level
affine Lie algebras, the GUT explanation for absence of fractional charges {\it does} work, but it is imposed by hand.
 In other words, in string theory
one cannot have it both ways: a natural explanation for GUTs, plus a natural explanation for charge quantization. Of course GUT coupling convergence and charge quantization are extrapolations of low energy empirical observations. Perhaps they are simply
wrong, and perhaps future experimental results reveal evidence for fractional charges, or counter-evidence for coupling unification.
But if we take these
empirical observations as real indications of fundamental physics, perhaps we ought to conclude that they point to an entirely different kind of theory, 
where unification is a fundamental feature, as it once seemed to be in the
heterotic string.

The common phenomenological answer to these objections is that we should merely try to find a string theory that fits the
data, no matter how rare or unnatural it may seem in the full set of possibilities. Indeed, finding an exact string theory realization
of the Standard Model would be tremendously important.  But of course such an achievement becomes less and less impressive as the known landscape becomes larger and larger. Furthermore there is a serious danger in dismissing apparent counter-evidence too quickly.
If
some feature -- such as the absence of fractional charges or the correct number of families -- is difficult to get, this fact may provide
essential information. It may indicate that one is considering the wrong class of string theories or that one is seeing artifacts
of an approximation.
It may also
indicate that we are missing some essential part of the underlying physics, and that the rare feature is enhanced by some mechanism
beyond current control, such as moduli stabilization. 

In certain cases, a rare feature may be needed for the existence of observers. 
In the context of the string landscape
it is nearly inevitable that the (poorly named) ``anthropic principle", in some form will play a r\^ole 
in determining the features we observe: other universes could be possible, and many of them would be lifeless. 
 The most conservative use of that principle is that one should not worry too much about
rare features if they have a potential anthropic explanation. Any such explanation necessarily consists of two parts:
an understanding of how a certain feature is distributed in the landscape, and an understanding of how changing it affects
the existence of life.  Both these issues can be studied scientifically. Nevertheless, the mere mention of the
word ``anthropic" tends to generate controversy. 
For a detailed discussion of these issues and the attitudes towards them we refer to \cite{Schellekens:2008kg}.
Of course, 
if all of the above fails we may ultimately have to conclude that we simply live in a universe that, for no known
reason, is rare in certain respects, but that is the least attractive option. 


In this paper we will focus on two issues: fractional charges and the number of families. Neither of these is totally immune
to anthropic considerations. Fractional charges can affect the evolution of matter in the universe and the chemistry and
nuclear physics upon which life is based in numerous ways. Perhaps their most important generic feature is that they
are stable (although in some cases they may be bound by non-Standard Model interactions into integer charge
bound states, which need not be absolutely stable, analogous to the proton). Stable additional matter will affect all abundances
of particles, will change big bang nucleosynthesis, the expansion of the universe and the energy balances of stars and nuclei, and will drastically alter chemistry, just to mention a few obvious consequences. In extreme cases, it may even be possible for the electron to decay into
two or more light fractionally charged particles. 
Even then, one is still facing the question whether or not  different kinds of life could be built
out of the fractional charges themselves. 
There is a bewildering number of possibilities for the masses and interactions of fractionally charged matter,
and it is a major challenge to come up with a convincing general argument against their existence, especially since 
additional fractional charges complicate physics substantially. 
But it seems implausible that the answer to the question: ``why haven't we
observed fractionally charged particles?" is: ``because no life is possible in their presence".  

Roughly the same can be said about
the number of families:  there is currently no argument why three families would be required for
life to exist.
Our kind of life requires at least one family, and if there are more than eight the strong coupling constant
becomes weaker at low energy, which is probably fatal for nuclear physics and life. Detailed examination of the running of all couplings, 
including higher loop effects,  may decrease this interval. The existence of a large top quark mass may turn out to be crucial, since it has
large loop effects on some couplings, and may affect weak symmetry breaking.  If true, this would increase the minimum number
of families to two. If the CKM CP-phase is somehow essential for life (perhaps through a presently less plausible
r\^ole in baryogenesis) the minimum would be three. 
In the absence of any anthropic explanation for three families, one may hope that the family
distribution in the landscape peaks at the value three. 
Otherwise
we may
just have to accept
that we ended up in a universe with three families just by chance. 
This would be still be perfectly acceptable 
if the family distribution is reasonably flat. 
%
%
However, previous studies of family number distributions seem to suggest that
in fact precisely the number three is difficult to obtain. 
In the area of orientifolds, systematic studies have been done for a variety of Standard Model features 
\cite{Dijkstra:2004cc,Gmeiner:2005vz,Douglas:2006xy,adks,Gmeiner:2008xq}. Although most 
of this work has focused on getting the Standard Model itself, a number of conclusions can be drawn about distributions
of certain parameters and discrete properties in its neighbourhood, such as coupling constants, the presence of various
kinds of GUTs, the number of families, or the presence of fractional charges. These studies
revealed a disturbing dip in the number of families, precisely for the observed value of 3, which occurred less frequently than
2 or 4 by two to three orders of magnitude \cite{Dijkstra:2004cc,Gmeiner:2005vz}.  

In order to appreciate the foregoing anthropic considerations, suppose we were discussing the number of colors $N_c$ rather than
the number of families. 
It is easy to define variants of the Standard Model with a different number of colors, with appropriate
changes in the electric charges of the quarks. In orientifold models, such variations have a natural quiver description similar
to the Madrid model, most easily obtained by starting with a Pati-Salam type configuration $(N_c\!+\!1,2,1)+(\overline{N_c\!+\!1},1,2)$ of
$SU(N_c+1)\times SU(2)\times SU(2)$.  In orientifold models 
the color distribution can just as easily be studied systematically as the family
distribution  (note that in heterotic models based on $SO(10)$-embeddings this is not possible). But for color, as opposed to the
number of families,
the numbers 1, 2 and 4 lead to entirely different physics than 3. There would be no baryons
at all, or only bosonic ones. A distribution of colors with a big dip at the value of three would not be a cause for concern, because
the neighbouring values can be ruled out on plausible anthropic grounds. The same cannot be said for the value $N_c=5$. However, it is
likely that any such distribution would drop off rapidly with the number of colors, since tadpole cancellation
requirements make it harder to accommodate larger stacks of branes. If that is indeed true, one can say that in the orientifold
landscape the observed value $N_c=3$ is understood by a combination of anthropic arguments and distributions.
In any case it is much easier to accept a dip at three in
the color distribution than a similar dip in the family distribution.

This immediately raises the question if
this dip in the number of families is an artifact of orientifolds or of the methods used (rational CFT or orbifolds), which is precisely one of the reasons for
studying the same question in heterotic strings. 
In that area, much less is known
so far. There is a fairly extensive amount of information on possible Hodge numbers or other topological data of Calabi-Yau manifolds ({\it e.g.} \cite{Kreuzer:2000xy},
 \cite{Gabella:2008id}) which provides family distributions for $E_6$ and other GUT gauge groups, 
 but there is little on family distributions for the Standard Model gauge group for exact string constructions. This is even more true for the presence or absence of fractional
charges in the massless spectrum. The only exception is the free fermionic construction, where results on distributions of gauge
groups are available 
\cite{Dienes:2006ut,Dienes:2007ms}, and where recently some studies with a similar focus as the present work have been made
 \cite{Faraggi:2006bc,Assel:2009xa,Rizos:2010sd}.
 
 The fact that three families can be obtained from heterotic strings has been demonstrated in
 various constructions; some early references are for example \cite{Greene:1986bm,Ibanez:1987sn,Bailin:1987xm,Casas:1988hb,Faraggi:1989ka}. In the context of Gepner models
 examples were found in \cite{Gepner:1987hi}, \cite{Schellekens:1989wx} and \cite{Blumenhagen:1996vu}.
There is also a large body of recent work on getting as close as possible to the currently known
Standard Model spectrum using heterotic strings (see {\it e.g.} 
\cite{Donagi:2004ia,Lebedev:2006kn,Blaszczyk:2009in,Anderson:2009mh,Faraggi:2010fi} 
and references therein), using combinations of string theory and
effective field theory. But this is not what we are aiming at. The question we are addressing here is not
if some of these features of the Standard Model can be realized (undoubtedly they can), but how rare they are. 

The work referred to in the previous  paragraph uses either geometric methods or free conformal field theory. We wish to explore
the area in between.
The class we consider,
interacting rational CFT, of which minimal N=2 tensor products are a special case, forms an interesting
area in between free field constructions and geometric constructions. The former allow the easy 
construction of many examples, but are limited in scope, since a free CFT is obviously a very special
case. The results of orientifold model building demonstrate that. Interacting CFTs provide a much richer
set of models, and much less constrained and more generic-looking distributions of physical quantities. 
To make a similar comparison for heterotic strings, we compare our results with free fermionic
constructions in an otherwise identical situation. Geometric methods are even broader in scope
than interacting CFTs, but it is harder to compute exact spectra.

This paper is the first in a series of studies of heterotic strings built using minimal $N=2$ model tensor products, thus providing
the missing heterotic counterpart to \cite{Dijkstra:2004cc} and \cite{adks}. In the present paper we take up the line of research
abandoned after \cite{Schellekens:1989wx}, and study $(0,2)$ models with broken $SO(10)$. These missing old results
will set the stage for future work, where we will also consider breaking of the $U(1)_{B-L}$ group, the extra heterotic $E_8$ 
factor and replace one or more of the minimal model building blocks in the bosonic sector. The class we consider is of course
only a subset of all possible heterotic strings, with the special property that the Standard Model is embedded in the canonical
way in $SO(10)$, and precisely that $SO(10)$ that is linked, via modular invariance, to the NSR model in the fermionic sector. 
This is the class of heterotic strings considered first in 1985 \cite{Candelas:1985en}
\cite{Witten:1985xc}, and that generated a lot of excitement because families
of $E_6$ and subgroups of $E_6$ came out automatically. Meanwhile even within the area of heterotic strings many other possibilities
have been explored. 

We will focus on two questions that remained unanswered two decades ago: Will the breaking of $SO(10)$ and/or left-moving world-sheet
supersymmetry allow us to change the somewhat disappointing old results regarding the number of families? With the exception of the three-family model found by Gepner, in all other cases the number
of families was found to be a multiple of 6 or 4 in 1989. Breaking some symmetries may affect that, and
indeed we will see that it does. The second question is:
How do the
fractionally charged particles appear in the spectrum? Can they be massive, or at least vector-like, and how often does that happen?  In 1989 only a few $SU(3)\times SU(2) \times U(1)$ spectra were examined, and they were discarded because fractional charges were present in the massless spectrum. But it is not excluded that
deeper searches provide examples where all of them are massive. Furthermore the chirality of the
fractional charge representations was not examined. Vector-like particles exist in essentially
all exact string spectra. Either they generically get a mass if the theory is perturbed and/or supersymmetry is broken, or we should
conclude that string theory is wrong. 
But if indeed they generically do get masses, the same logic can be
applied to vector-like fractionally charged particles.

This paper is organized as follows. In the next section we discuss and review the CFT model building aspects, define the
class we are examining and explain
how we scan the set of MIPFs. In section
3 we discuss charge quantization, and display a range of possibilities in between $SO(10)$ unification and complete
$SO(10)$ breaking, and show how at least half-integer charges or third integral charges can be avoided without paying a price.
In section 4 we discuss our search results. We present the relative frequencies of occurrence of the various
possibilities for charge quantization, and present an intriguing observation regarding the quantization of the number
of families, which appears to be valid for other CFT constructions (in particular Kazama-Suzuki  \cite{Kazama:1988qp} models) as well. We also
present some distributions of non-chiral quantities (singlets and mirrors) for the different spectra we have found.
In section 6 we present some results on the only tensor product
that gives rise to three families, based on the original example found by Gepner \cite{Gepner:1987hi}.  
In the last section we formulate our conclusions.

\section{CFT construction}

The class of heterotic strings considered here is defined as follows. We will make use of the bosonic string map \cite{LLS},
and we first build a four-dimensional bosonic string theory with the special property that its right-moving sector can be mapped
to a fermionic sector. This bosonic string is built as a tensor product of
a set of CFTs with a total central charge equal to 5, an $E_8$, level 1 affine Lie algebra,
and a $c=9, N=2$ internal CFT. The latter CFT is
obtained by tensoring certain building blocks, which can be any $N=2$ CFT. The $c=5$ system, denoted ${\cal S}$ in the following,
 is required to allow a chiral
algebra extension to $D_5$. We start with the diagonal partition function of this entire system. 

Now we extend the chiral algebra of the right-moving sector in the following way. First we extend the $c=5$ system to $D_5$.  
All tensor products are ordinary,
non-supersymmetric CFT tensor products, which in particular means that the set of states contains combinations of NS and R
states, which violate world-sheet supersymmetry.
We project all of these out by extending the right chiral algebra by all combinations of the vector of $D_5$ with the world sheet
super currents of the $N=2$ building blocks (sometimes called ``alignment currents"). 
Next we extend the right chiral algebra by a combination of the spinor of
$SO(10)$ and Ramond ground states of each of the $N=2$ building blocks. 
Finally we replace, in the right-moving sector, the characters of
$SO(10)  \times E_8$ by Lorentz-covariant NSR characters, and we obtain a heterotic string with $N=1$ space-time supersymmetry. 

All the foregoing extensions are required in the right sector only. However, if we keep the left sector unchanged the result
would not be modular invariant. The latter can be preserved by making exactly the same extensions in the left sector as well,
but this automatically implies the presence of phenomenologically undesirable gauge symmetries. 
In the class of models considered here, the Standard Model gauge bosons originate from spin-1 currents in the
CFT ${\cal S}$. The chiral algebra extensions lead to the extension of ${\cal S}$ to $SO(10)$ and finally $E_6$.

However, the left and right algebras only need to be identical from the perspective of the modular group.  This
allows in particular the replacement, in the left sector, of spin-1 currents by higher spin currents, all with the same mutual monodromies. 
The alignment
currents have spin 2 and have no phenomenological implications for the gauge group, and hence there is no need
to replace them, but requiring them to be identical in the left and right sector is an unnecessary restriction, which we might as well 
remove also. If only the space-time supercurrent is realized asymmetrically, a $(1,2)$ model is obtained. If the
extension to $SO(10)$ or world-sheet supersymmetry is realized asymmetrically we will call the result a  $(0,2)$ model 
\cite{Silverstein:1995re,Kachru:1995em,Distler:1995bc,Blumenhagen:1995tt,Blumenhagen:1995ew,Blumenhagen:1996xb,Blumenhagen:1996gz,Berglund:1995dv,Kreuzer:2009ha}.

In general, very little is known about asymmetric partition functions. They can be constructed systematically using simple
currents (including free bosons as a special case) or free fermions. Furthermore a few exceptional examples are known,
such as the meromorphic $c=24$ CFTs \cite{Schellekens:1992db} (which can be freely combined in the left and right sectors). A more radical 
approach is to replace the conformal building blocks themselves by distinct, but isomorphic ones.  An example are
 the ``lifted" Gepner models \cite{GatoRivera:2009yt}. 
 
 In this paper we keep the same building blocks, namely $N=2$ minimal models, in both sectors. 
  Essentially we will do precisely what was announced in \cite{Schellekens:1989wx}, but was only
 partly completed then. The novel features of the present paper in comparison to \cite{Schellekens:1989wx} are
 \begin{itemize}
 \item{$SO(10)$  breaking. This was only mentioned in \cite{Schellekens:1989wx}, but not carried out because of
 computational limitations.}
 \item{Asymmetric realizations of alignment symmetries, yielding $(0,2)$ spectra. In \cite{Schellekens:1989wx} this was only done for
 a single tensor product, the $(1,16^*,16^*,16^*)$}. In all other cases only $(2,2)$ and $(1,2)$ spectra were considered.
 \item{Use of the general simple current construction described above. This
 formalism was not yet available in 1989. 
 In \cite{Schellekens:1989wx} only products of several single current MIPFs were considered, a limitation which
 misses some cases with discrete torsion.}
 \end{itemize}

To build MIPFs we make use of the formalism developed in \cite{GatoRivera:1991ru,Kreuzer:1993tf}, which
 can be shown to generate the most general MIPFs with non-vanishing off-diagonal multiplicities $Z_{ij}$ that lie entirely 
 on simple current orbits: $Z_{ij} \not = 0$ only if $i$ and $j$ are on the same simple current orbit. A general simple current 
invariant is described by a set of simple currents forming a discrete
abelian group ${\cal G}$, and
a matrix $X$. The group ${\cal G}$ is a product of $k$ cyclic factors, each
generated by a current $J_s$. The relative monodromies of these currents define a monodromy matrix $R$ that is defined as 
$R_{st}=Q_s(J_t) \mod 1$, 
plus a further constraint that fixes its diagonal elements modulo 2, depending
on the conformal weight of the currents.
 The matrix $X$ (defined modulo 1) has to
satisfy
the equation
\beq\label{Xeq} X+X^T=R .\eeq
The matrix $X$ determines the matrix $Z({\cal G},X)_{ij}$ in the following way: 
$Z_{ij}$ is equal to the number of solutions $J$ to the 
conditions
\begin{eqnarray} j&=&Ji, J \in {\cal G}\ \ \ \ \  \hbox{~and~}\cr  Q_K(i)+X(K,J)&=&0 \mod 1  \hbox{~~for all~}
K\in {\cal G}\ .
\end{eqnarray} 
Here $X(K,J)$ is defined by writing $K$ and $J$
in terms of the generating currents $J_s$, $J=(J_1)^{n_1}...(J_k)^{n_k}$,
$K=(J_1)^{m_1}...(J_k)^{m_k}$. Then
$$ X(K,J) \equiv \sum_{s,t} n_s m_t X_{st}\ . $$
The group ${\cal G}$ can be any subgroup of the full simple current group. The matrix $X$ is a set of rational numbers. For
each tensor product this yields a huge, but finite set of possibilities. For example, an $N=2$ minimal model 
at level $k$ has a simple current $\Zbf_{2k+4}\times \Zbf_2$ for $k$ even and $\Zbf_{4k+8}$ for $k$ odd. For the $SO(10)$
subgroup ${\cal S}$ we will use $SU(3) \times SU(2) \times U_{30} \times U_{20}$ in this paper (see the next chapter). 
If we choose the internal sector $(3,3,3,3,3)$ (which denotes the tensor product of five $k=3$ minimal models), the total
simple current group is $\Zbf_3\times \Zbf_2\times \Zbf_{30}\times \Zbf_{20} \times (\Zbf_{20})^5$. The total number of MIPFs can
be computed by splitting this group into prime factors, and multiplying the number of MIPFs corresponding to each prime factor. There exists
a formula for the number of MIPFs for discrete groups $(\Zbf_p)^K$, $p$ prime \cite{GatoRivera:1991ru}:
$$ N_{\rm MIPF} = \prod_{\ell=0}^{K-1} (1+p^{\ell}) $$
The generalization of this formula to factors $Z_{p^n}$, $n>1$, is not known. In this example, the seven factors $\Zbf_5$ 
 already contribute 1.202.088.011.709.312, the two factors $\Zbf_3$ multiply this by eight, and then there still is a 
 factor $(\Zbf_2)^2 \times ( \Zbf_4)^6$ which multiplies this with another huge number. These numbers are reduced by
 the requirement of the presence of $SO(10)$, alignment and supersymmetry currents in the fermionic sector. Furthermore, in this
 particular example there will be a redundancy due to the permutation symmetry of the five $k=3$ models.

The building blocks of the internal CFT that we will use are the $N=2$ minimal models. They can be combined in 168 ways to form $c=9$ CFTs.
In addition we consider the 59 CFTs obtained by using exceptional $SU(2)$ modular invariants, which exist for
level 10, 16 and 28. 

Finally, to put the results in perspective, we also consider $N=2$ building blocks constructed
out of free fermions.  Here the $c=9$ internal sector is build out of combinations  of $N_c$ complex and $N_r$ real fermions
($(N_c,N_r)$ can be (9,0), (7,4), (5,8) and $(3,12)$) with world-sheet supersymmetry realized through the
so-called ``triplet constraint" \cite{Antoniadis:1985az,Kawai:1986va,Antoniadis:1986rn}. The symmetric $(2,2)$ combinations
of these models were considered in  \cite{Kiritsis:2008mu} as a starting point for orientifold constructions. In total,
one may distinguish 62 different diagonal CFTs, which differ in the way the real and complex fermions are combined
into triplets.
This is however not the most general free fermionic construction: we only build the $c=9$ internal CFT out of free fermions
(we leave $E_8$ intact), and
we exclude extensions by combinations of spinor currents of real fermions, because they are not simple currents.
This is also the reason \cite{Kiritsis:2008mu} why we must keep three fermions complex: otherwise we cannot impose space-time supersymmetry as
a simple current invariant\rlap.\footnote{Note that there is in principle no obstacle for using exceptional MIPFs in heterotic
constructions of this kind. One can simply multiply the modular matrices. This is how we deal with minimal model
exceptional invariants. By contrast, in orientifold constructions exceptional MIPFs are an obstacle, because the boundary state formalisms
requires fixed point resolution. The main problem in asymmetric heterotic strings is that there is no easy way to use exceptional
invariants asymmetrically. If one wishes to do that for spinor currents of real fermions, it would be better to use directly the complete free fermionic formalism developed
in \cite{Kawai:1986va,Antoniadis:1986rn}. Here we merely use a large subset for comparison with Gepner models.}
 One may 
also combine Gepner and free fermionic building blocks, but we will not consider that option here.

\subsection{Search procedure}

 It seems essentially impossible to search the huge set of MIPFs systematically, even if one takes the constraints into account from the
 start. In the present paper we deal with this as follows. First we determine in the fermionic sector all character combinations that
 contribute to the massless sector, and that are local with respect to the $SO(10)$ currents, the alignment currents and the
 space-time supersymmetry current. Then we generate a subgroup ${\cal G}$ by choosing at random $M$ elements of the 
 full simple current group. Then we choose at random a matrix $X$ that satisfies (\ref{Xeq}). We compute the matrix $Z({\cal G},X)_{ij}$,
 act with it on the set of right-moving massless states, and read off the set of left-moving massless states with which they are
 combined.

\subsection{Mirror symmetry}

The complete set of MIPFs we consider here has an exact mirror symmetry. This is because we can always
use a simple current to map, for any given spectrum,  spinors of $SO(10)$ to their conjugates. Since we take random
samples of the set of MIPFs, mirror symmetry is not automatic in the results, but will be approached asymptotically for
large samples. It can therefore be used to get some idea about the level of saturation of the search. Another way
of detecting that is to consider the frequency of occurrence of distinct MIPFs. If there are many that have occurred just once,
it is likely that many have not been seen yet. These two ways of checking saturation do indeed agree: if the minimal occurrence
frequency is larger than five, we find in all cases that mirror symmetry is exact. 
This notion of mirror symmetry should not be confused with the one used in the context of Calabi-Yau compactification. What
we are talking about is a trivial transformation of the partition function, which always exists, because we allow the
transformation to act on the NSR part of the theory. It is not trivial that a mirror map exists that acts only on the internal
degrees of freedom. Presumably the arguments presented in \cite{Blumenhagen:1996vu} imply that it does, but we have
not investigated this. We merely use mirror symmetry here as a bookkeeping tool.

\section{Charge Quantization}

In the heterotic string with the $SU(3)$ and $SU(2)$ factors realized as affine Lie algebras
at level 1, one gets conventional $SU(5)$-type coupling constant convergence if the $U(1)$ factor
is $U_{30}$: the free bosons compactified on a circle with radius such that the number of primaries
is $30$. The $Y$ charge of the quark doublet corresponds to the smallest $U_{30}$ charge. 
%

The canonical way of obtaining the Standard Model from heterotic strings is to embed
it in the level 1 $SO(10)$ factor that appears automatically in (2,2)-type constructions. 
Then $SO(10)$ is decomposed to $SU(3) \times SU(2) \times U_{30} \times U_{20}$.
To build a heterotic string theory using the bosonic string map explained in the previous sections, 
we start with a bosonic string with this subgroup structure symmetrically in the
left- and the right-moving sector. Then we
extend this subgroup in the right (fermionic)
sector back  to $SO(10)$, and then we map the $SO(10) \times E_8$ characters to the NSR
model. 

The extension of $SU(3) \times SU(2) \times U_{30} \times U_{20}$ that yields $SO(10)$ is of order 30, and
is generated by the simple current $(3,2,1,4)$ (representations are denoted as $(\hbox{dim},\hbox{dim},p,q)$, where $p$ and $q$ are
the $U_{30}$ and $U_{20}$ charges). In the bosonic sector we may encounter 
any cyclic subgroup of this extension, generated by powers of $(3,2,1,4)$ that are divisors 
of 30. The possibilities are listed in Table 1. 

Each of the eight possibilities gives rise to
a subgroup of $SO(10)$ which restricts the allowed Y-charges. But the Y-charges may
also be restricted by higher spin currents in the chiral algebra, that do not lead
to massless vector bosons in the spectrum. In the sixth column we indicate the
charge quantization imposed by the gauge group, and in the seventh column the charge
quantization imposed by the full CFT. Modular invariance implies that this also works in 
the other direction: the absence of certain fractional charges implies the presence of certain currents
in the chiral algebra, by the same arguments as those used in \cite{Schellekens:1989qb}. 
In particular, the absence of half-integer charges implies the presence
of the $10^{\rm th}$ power of the generator, whereas the absence of third-integral charges
implies the presence of the $15^{\rm th}$ power. Absence of both half-integer and third-integer charges
(and hence absence of all fractional charges) implies both the $10^{\rm th}$ and the $15^{\rm th}$
power to be present, and hence by closure of the algebra also the $5^{\rm th}$ power. This is
the root of $SU(5)$.  Unlike the $5^{\rm th}$ power, the $10^{\rm th}$ and the $15^{\rm th}$ are
currents of conformal weight larger than 1, and hence their presence does not lead to massless
vector bosons. 

So we see that in this class of heterotic strings there is a very interesting, and essentially stringy,
mechanism available to prevent some of the fractional charges in the spectrum without having Grand Unification
of gauge symmetries. This mechanism can work to prevent either half-integer charges or third-integral charges,
but not both: if both constraints are imposed the $SU(5)$ roots are automatically present as well.
In a theory with one of these constraints imposed (but not both)
Standard Model gauge couplings converge at the string scale, but this would not imply an enlargement of the gauge 
symmetry. 
One can make the same mechanism work for all fractional charges but  at the price of considering higher level
affine algebras. In general, if one considers the gauge group
$$ SU(3)_{k_3} \times SU(2)_{k_2} \times U_{2 k_1}\ , $$
with a quark doublet realized as $(3,2,1)$, 
it is possible to project out all fractionally charged states provided \cite{Schellekens:1989qb}
$$ k_1+9k_2 + 12 k_3 = 0 \mod 36$$
and one gets standard coupling unification if $k_2=k_3$ and $k_1=15k_2$. This has a solution for any integer $k_2$, and in all cases except $k_2=1$ that solution involves higher spin currents only. Unfortunately, there is a canonical
construction -- the one we use here -- only for $k_2=1$, and furthermore higher values of $k_2$ allow other unwanted states:
higher tensor representations of $SU(3)$ and $SU(2)$. To our knowledge such higher level models have not been
constructed (they should not be confused with higher level GUT models, which {\it have} been constructed (see {\it e.g.} \cite{Lewellen:1989qe,Font:1990uw,Chaudhuri:1994cd,Kakushadze:1996tm}) in the class described
here there is no GUT gauge group). 

It is instructive to compare with orientifold models (without any intention to suggest exact dualities).
The fundamental problem with charge quantization in heterotic strings is
that there is no {\it a priori} correlation between color and the Standard Model $U(1)$. Both come from different Lie algebra factors. Grand Unification
imposes the required correlation at the price of introducing extra massless gauge bosons. Orientifold models force us to
impose such a correlation from the start. The reason is that we must get the quark doublet $(3,2,\frac16)$ from a bifundamental, and
hence inevitably the $U(1)$ charge must come from either the $U(3)$ stack or the $U(2)$ stack. This makes a correlation
between color and charge arise in a much more natural way. 
On the other hand, in orientifold models it is less clear
why one family should have the structure we observe, whereas in heterotic strings this follows from the GUT embedding. However,
in both cases we are asking the same question: if we require the existence of standard quarks and leptons in the spectrum,
does this imply or allow the existence of other $SU(3)\times SU(2)\times U(1)$ representations that do not belong in
a Standard Model family?  

In orientifold models, imposing the existence of the $(3,2,\frac16)$ representation requires the $Y$ charge to receive
the following contributions from the $U(1)$ generators $Q_{\bf a}$ and $Q_{\bf b}$ of the $U(3)$ and $U(2)$ stack
$$ Y= (x-\frac13) Q_{\bf a} + (x-\frac12) Q_{\bf b}  + \ldots \ ,$$
Here the quark doublet is assumed to be a $(V,V^*)$ bifundamental. The parameter $x$ is determined by the requirement
that all charged particles in a family can be obtained. 
Under very general conditions (spelled out in \cite{adks})
the following possibilities exist. If all Standard Model particles are realized using orientable configurations, no constraint
on $x$ is obtained. In that case, all Standard Model particles are $(V,V^*)$ bi-fundamentals, and if each brane stack ${\bf s}$ contributes
$x Q_{\bf s}$ to $Y$, all these contributions cancel. If one of the Standard Model particles originates from a $(V,V)$ bi-fundamental or
a rank-2 tensor, then $x$ is either 0 or $\frac12$. We then get the following possibilities for fractional charges:
\begin{itemize}
\item{$x=0$. Then all states in the perturbative spectrum respect Standard Model charge quantization. This class includes $SU(5)$ GUT.}
\item{$x=\frac12$. Then all strings between the observable sector branes respect charge quantization, but open strings
between hidden sector and observable sector branes lead to half-integer charge particles. If there is no hidden sector,
Standard Model charge quantization holds. This class includes the Pati-Salam model $SU(4)\times SU(2)\times SU(2)$.}
\item{Any other value of $x$ (although $x$ is usually constrained by the requirement that $Y$ does not acquire a mass from axion mixing). Then Standard Model charge quantization holds if the entire observable sector brane configuration is orientable, and
there is no hidden sector. Non-orientable matter introduces fractional charges $\pm 2 x$, observable-hidden matter gives rise
to fractional charges $\pm x$. This class includes the trinification model $SU(3)^3$ for $x=\frac13$. This value is special since
there is no contribution to $Y$ from the color stack, and the existence of this configuration is not constrained by the $Y$-axion mixing condition.}
\end{itemize}
The first class has some similarity with the $k_2 > 1$ models described above. Full $SU(5)$  GUT unification
is possible by having a stack of five unitary branes, but if the five branes are split in groups of two
and three one gets something akin to the higher level $SU(3)\times SU(2) \times U(1)$ models
described above.  Using orientifolds one gets simple and natural realizations  of half-integer and third-integer charges, just
as with heterotic strings. However, there is no natural orientifold analog of the sixth-integral fractional charges that occur
abundantly in heterotic strings.

Let us return to heterotic strings with $k_2=1$, the case of our interest. 
Consider the various subgroups listed in table \ref{TableG}.
In order to project out a subset of the
orbit, it must be possible to replace some of the currents by others of the same order. This is a necessary condition
for being able to realize the extension asymmetrically. 
To see the options it is sufficient to consider the prime order generators, which are of
order 2, 3 or 5. 

If the internal CFT does not contain a simple current of order 5, the fifth order current
$(1,1,6,4)$ cannot be projected out. The only current in the left-moving sector that can replace
it is then its own conjugate, which is of no interest. The same is true for the order 3 current 
$(3,1,10,0)$. Furthermore, in both cases the presence of an order 5 (respectively 3) current
of non-integer conformal weight in the internal sector is sufficient to allow
the $SO(10)$ current $(1,1,6,4)$ (respectively $(3,1,10,0)$) to be projected out. One can always combine
a suitable power of that current with the order 5 or 3 current in $U_{30}$ in order to get an
integer spin current that is non-local with respect to the $SO(10)$ currents one wants to project out. This
would not work for integral weight order 5 or 3 currents in the internal sector. Although they
would seem perfect candidates for replacing the $SO(10)$ currents, they are local with respect to them,
and hence cannot affect them. Fractional spin currents of order 5 or 3 are present in $N=2$ minimal models
if $k+2$ contains a single factor 5 or 3, because the simple current group  is 
$\Zbf_{2k+4}\times \Zbf_2$ for $k$ even and $\Zbf_{4k+8}$ for $k$ odd.

For the order two current $(1,2,15,0)$ the discussion is a bit more
complicated, because the building blocks we use always contain currents of even order. In a suitable extension,
these can become currents of order two that could replace $(1,2,15,0)$ in the bosonic sector. However, one must
also find left images of the alignment currents, which are also of order two. This can all be analysed
in detail, but it is easier to find out empirically. The result is that the order two current can be
projected out in all but two cases with exceptional MIPFs. The reason exceptional MIPFs can 
behave differently is that they automatically imply a D-type extension by a current of order two,
which adds an extra constraint to the set of order two currents.

The different possibilities can be summarized as follows:
\begin{itemize}
\item{The currents of order 2, 3 and 5 can all be projected out. Then all groups are possible. This happens for example
for the tensor product $(3,4,13,13)$, which contains currents of order 2, 3 and 5.}
\item{The current of order 5 cannot be projected out. Hence the order in column 4 of table \ref{TableG} must be divisible by 5. The only allowed groups are then Nrs. 3, 5, 6 and 7 (LR, Pati-Salam or $SO(10)$). An example is the tensor product $(1,1,1,1,1,1,1,1,1)$. In these models $SU(2)_R$ cannot be broken. }
\item{The current of order 3 cannot be projected out. The only allowed groups are then Nrs. 2, 4, 6 and 7 in table \ref{TableG}.
An example is $(3,3,3,3,3)$. In these models there can only be half-integer and integer charges.}
\item{The currents of order 3 and 5 cannot be projected out. The only allowed groups are then Nrs. 6 and 7. Examples are $(6,6,6,6)$ and all
free fermionic models. The only allowed groups are Pati-Salam or $SO(10)$, and only integer and half-integer charges are possible. }
\item{The currents of order 2 and 5 cannot be projected out. The only allowed groups are then Nrs. 5 and 7. This happens only for the
exceptional models $(1,16,16^*,16^*)$ and   $(1,16^*,16^*,16^*)$. Only integer and third-integer charges can occur, not half-integer and sixth-integer. These are the only cases for which the options for charge quantization cannot be read off directly from the
$k$-values of the factors, and also the only ones where the order two current cannot be projected out. }
\end{itemize}

\renewcommand{\arraystretch}{1.2}
{\small
\begin{table}[P]
\begin{center}
\vskip .7truecm
\begin{tabular}{|c|c|c||c|c|c|c|}
\hline
\hline
Nr. & Name & Current & Order & Gauge group  & Grp. & CFT \\
\hline
0 &\small SM, Q=1/6& \small $(1,1,0,0)$ & $1$ & {\footnotesize  $SU(3)\times SU(2) \times U(1) \times U(1)$} &  $ \frac16$  & $ \frac16$  \\
1 &\small SM, Q=1/3 &\small $(1,2,15,0)$ & $2$ &  {\footnotesize $SU(3)\times SU(2)\times U(1)\times U(1)$}  &  $ \frac16$ & $ \frac13$  \\
2 &\small SM, Q=1/2 &\small $(3,1,10,0)$& $3$ &  {\footnotesize $SU(3)\times SU(2)\times U(1)\times U(1)$}  &  $ \frac16$  & $ \frac12$  \\
3 &\small LR, Q=1/6 &\small $(1,1,6,4)$ & $5$ & {\footnotesize $SU(3)\times SU(2)_L \times SU(2)_R\times U(1)$}  &  $\frac16$  & $ \frac16$  \\
4 &\small SU(5) GUT&\small $(\bar 3,2,5,0)$ & $6$ & {\footnotesize  $SU(5) \times U(1)$}  &  1  & 1  \\
5 &\small LR, Q=1/3 &\small $(1,2,3,-8)$ & $10$ &  {\footnotesize $SU(3)\times SU(2)_L \times SU(2)_R\times U(1)$}  &  $\frac16$  & $ \frac13$  \\
6 &\small Pati-Salam &\small $(\bar 3,0,2,8)$ & $15$ & {\footnotesize  $SU(4)\times SU(2)_L \times SU(2)_R$}  &  $\frac12$  & $ \frac12$  \\
7 &\small SO(10) GUT &\small $(3,2,1,4)$ & $30$ & {\footnotesize $SO(10)$} &  1 & 1 \\
 \hline
\end{tabular}
\vskip .7truecm
\caption{\small List of all Standard Model extensions within $SO(10)$ and the resulting group theory and
CFT charge quantization (last two columns). We refer to these subgroups either by the label in column 1 or by the name in column 2, where
``LR" stands for left-right symmetric.}
\label{TableG}
\end{center}
\end{table}
}

\section{Results}

\subsection{Spectrum selection}

The spectra we have obtained have a gauge group containing $SU(3)\times SU(2)\times U(1)$ as a subgroup. 
We assume that
eventually all other gauge symmetries are broken to this subgroup, so that only chirality with respect to this subgroup matters; this
is always what we mean by chirality henceforth.
If a spectrum contains chiral matter that does not fit in a Standard Model family, then
there is often not even a sensible definition of the number of families, because 
the net number of chiral representations for the usual quark and lepton multiplets Q, U, D, L or E can be
different; if on the other hand only the representations Q, U, D, L and E are present, anomaly cancellation
guarantees that they have the same chiral multiplicities. Anomaly cancellation is exact in all the
spectra we obtain, and in particular there are no $U(1)$'s whose anomalies are cancelled
by the Green-Schwarz mechanism. This is because there is an unbroken $E_8$ factor in the gauge
group, which cannot contribute to  a ${\rm Tr}  F\  {\rm Tr} F^2$ term in  the anomaly polynomial. 
But on the other hand modular invariance guarantees that  the ${\rm Tr} F^2$  factor must have contributions from all gauge groups, including the unbroken $E_8$. 

Spectra with chiral exotics are just counted, but not distinguished. All remaining spectra are
distinguished on the basis of the following data: 
\begin{itemize}
\item{The combination of minimal models and the choice of exceptional MIPFs.}
\item{The number of families.}
\item{The number of Q, U, D, L, E mirror pairs, counted separately.}
\item{The number of singlets.}
\item{The total number of fractionally charged $SU(3)\times SU(2) \times U(1)$ representations.}
\item{The $SO(10)$ part of the gauge group (the eight possibilities listed in table \ref{TableG} below).}
\item{The total dimension of the gauge group.}
\item{The actual charge quantization observed in the massless sector.  This is usually precisely the CFT
quantization in Table 1, but in very rare cases it may happen that the smallest allowed charges are only
seen in the massive spectrum, and are absent from the massless spectrum.}
\end{itemize}
We ignore $B-L$ quantum numbers and treat right-handed neutrinos just like other Standard Model singlets. 

For each distinct spectrum, we keep track of the number of times it occurred in a random search 
as defined in  section 2.1. This leads to two ways of counting spectra: the total number of times
it occurred, and the total number of distinct ones. We will refer to these as occurrence counting
and spectrum counting. 
Only the former method is available for comparing
exotic and non-exotic spectra, since for the exotic ones we only have a total count.   This is therefore
the method used in the next subsection. For the other results we have a choice. For example, one
can make a distribution of the total number of singlets based on total occurrence, or plot
the number of distinct spectra with a certain number of singlets. The first method favors the
GUT spectra that are obtained with the least amount of simple current modification. 
Especially the simplest $(2,2)$ $E_6$ spectrum obtained using the diagonal invariant tends to come
out very frequently. Thus unmodified GUT spectra are overcounted relative
to the others.  In other words, the objects closest to the lamppost are most easily seen, but that does not mean
that they are more numerous.

True landscape statistics would require counting of distinct, moduli stabilized heterotic ground states,
and this is at present not possible. 
All we can reasonably expect to do now is to get some idea 
about rare and fairly generic features. We expect spectrum counting to be closer to actual
distinct vacuum counting than occurrence counting, and hence when possible we favor the former 
method. Note that spectrum counting leads to distributions that do not depend on the
choice of randomization. 

For all quantities except
exotics we have chosen the
latter method. To illustrate the difference between the two counting methods we show
the family distribution for the $(1,16^*,16^*,16^*)$ model in both ways in section 5.

\subsection{Occurrence of fractional charges}

\begin{table}[P]
\begin{center}
\vskip .7truecm
\begin{tabular}{|c||c|l||c|c|c|c|}
\hline
\hline
 & \multicolumn{2}{|c||}{Fractional charges} & \multicolumn{3}{|c|}{CFT type} \\
\hline
Type   &  Massless & Chiral? & Gepner  & Gepner(exc.) & Free Fermion\\
\hline
SM, Q=1/2 &   $\frac12$  & NO & $0.97\%$ & $0.30\%$  &   - \\
SM, Q=1/3 &   $\frac13$  & NO & $0.42\%$    &  $0.76\%$        &- \\
SM, Q=1/6 &   $\frac13$ & NO & $0.000023\%$   &   -       & -\\
SM, Q=1/6 &   $\frac16$ & NO & $0.23\%$   &     $0.28\%$     &- \\
LR, Q=1/3 &    NO   &   NO & $0.0000031\%$ & - & - \\
LR, Q=1/3&   $ \frac13$  & NO & $3.18\%$    &  $10.15\%$        &- \\
LR, Q=1/6 &   $\frac13$  & NO  & $0.0013\%$   &   $0.010\%$       &- \\
LR, Q=1/6 &   $\frac12$  & NO  & $0.000029\%$   &   $0.0000126\%$       & -\\
LR, Q=1/6 &   $\frac16$  & NO  & $1.06\%$   &  $2.86\%$        &- \\
Pati-Salam &  NO & NO  & $0.001\%$   &  $0.016\%$        &  $8.43\%$ \\
Pati-Salam &   $ \frac12$ & NO  & $15.53\%$   &  $9.41\%$        & $ 68.42\%$ \\
\hline
SU(5) GUT &  NO                 &  NO & $13.06\%$    &  $4.67\%$        &- \\
SO(10) GUT &  NO &NO  & $32.7\%$   &  $32.99\%$        & $21.63\%$ \\
SM, Q=1/2 &   $\frac12$  & YES & $1.09\%$    &  $0.09\%$        &- \\
SM, Q=1/3 &  $\frac13$  & YES& $1.63\%$    &  $0.82\%$        &- \\
SM, Q=1/6 &   $\frac16$ & YES & $0.66\%$   &   $0.25\%$       &- \\
LR, Q=1/6 &   $\frac16$  & YES  & $4.98\%$   &   $2.86\%$       &- \\
LR, Q=1/3&  $ \frac13$  & YES & $22.88\%$    &   $33.89\%$       & -\\
Pati-Salam &   $ \frac12$ & YES  & $1.65\%$   &   $0.78\%$       & $1.5\%$ \\
\hline
\end{tabular}
\vskip .7truecm
\caption{Allowed and observed fractional charges. In column 1 we display the CFT quantization as
shown in Table 1. In column 2 we show the fractional charges actually seen in the  massless sector. In 
column 3 we indicate if those fractional charges  are chiral. The other three columns show the relative frequencies of these cases in standard Gepner models, exceptional Gepner models and free fermion models. Spectra below the horizontal line are phenomenologically unacceptable.  }
\label{Fracs}
\end{center}
\end{table}

In table \ref{Fracs} we list the various possibilities and their relative frequencies. We distinguish
standard Gepner models and exceptional Gepner models, as well as free fermions. In each case
we have selected a random set of simple currents and a random choice of discrete torsion
parameters. For the Gepner models, we have considered simple current groups generated by
up to four currents, whereas for free fermions we have allowed up to six currents. 
This table is based on about $64 \times 10^6$ samples for standard Gepner models,
$8\times 10^6$ for exceptional ones and $130 \times 10^6$ for free fermions. 
In 
 the last three columns we indicate which percentage of these samples belonged to a 
 certain type of fractional charge realization.  Obviously there are many ways of distributing random numbers over
 the space of possibilities, so these numbers just give an idea of how rare
some of them are, and should not be confused with proper landscape statistics. 

The lines in the table are ordered according to desirability. Gauge groups closest to the Standard Model
appear first, and cases with chiral fractional charges are at the bottom. We regard anything below the line as unacceptable. Unbroken $SU(5)$ and $SO(10)$ GUTs are unacceptable because
their massless spectra contain no Higgses to break the GUT gauge group. One might still consider
composite Higgses, but this is too far-fetched to consider seriously. The other extended gauge symmetries, Pati-Salam and $SU(2)_R$, can in principle be broken by Higgses that are not
forbidden in the massless spectrum, and that in general do indeed occur. 

Above the line there are five cases where some {\it a priori} allowed fractional  charges are entirely absent from the 
massless spectrum (of course they {\it must} occur in the massive spectrum). In two of those there are no fractional
charges whatsoever, and in the other three the fractional charge quantum is larger than expected on the basis
of the algebra ({\it i.e} the CFT quantization listed in column 1 does not match the actual quantization for massless particles listed in column 2)).
All of these are
extremely rare in Gepner models. Furthermore, we have seen them only in purely non-chiral
spectra. In free fermionic models, the only possible gauge groups are $SO(10)$ and
Pati-Salam. In the latter case, we found that in about $8\%$ of all cases the expected half-integer
charges were completely absent from the massless spectrum. However, most of these spectra
are purely non-chiral as well; only about $1\%$ of them ($.071\%$ of the total) had chiral families
(either 12 or 24).  A similar  result was reported in  \cite{Assel:2009xa}. These authors looked
at a different class of free fermionic models, designed for getting three families, and also
reported a number of cases of Pati-Salam models without massless fractionally charged exotics. 

Even within the class of free fermionic models, the examples without massless fractional
charges, but with a chiral spectrum are rare. In our opinion, the reason we do not observe
fractional charges  is {\it not} that our universe ended up in such a rare ground state
just by chance. Within
the heterotic landscape, a more plausible explanation is that we ended up in a universe
with, to first approximation, vector-like fractional charges, and that these have the same fate as
other vector-like particles, namely acquire a mass. In other words, there is no good reason to prefer
spectra without massless fractional charges over those with vector-like fractional charges at our present level
of approximation, but it is reasonable to reject those with {\it chiral} fractional charges.
If we discard the $SU(5)$ and $SO(10)$ GUTs, 
the relative probability
for encountering only vector-like fractional charges is substantial (of order $40\%$ with our method of random sampling, but in any case not absurdly rare). 
But even this option is not free of problems. The fractionally charged particles must become
sufficiently massive and consequently rare to have escaped detection so far. 

The appearance of chiral fractional charges correlates almost perfectly with the appearance of chiral families: tensor
combinations in which the number of chiral families is always zero never had chiral fractional charges either. 
There is one tensor combination with chiral spectra, but without chiral exotics in any of those spectra, namely 
$(5,5,5,12)$. There is also precisely one free fermion model with that property, namely the one built out of 12 Ising
models and 3 free bosons, {\it i.e} the maximal number of real fermions we allow.

\subsection{The number of families}

In previous work \cite{Schellekens:1989wx} it has been observed that the number 
of families for the different MIPFs of any given Gepner model is quantized in certain units.
In that paper, for each model all simple current MIPFs were considered that gave rise to (2,2)
and (1,2) compactifications. The greatest common divisor of the number of families
for a given Gepner combination will be called the ``family quantum" $\Delta$. 
The following values for $\Delta$ were found (each exceptional modular invariant is treated as a
separate category):
$$ 120, 96, 72, 60, 48, 40, 36, 32, 24, 12, 8, 6, 4,  0 $$
plus the value 3, which only occurred for the tensor product $(1,16^*,16^*,16^*)$, where
the $*$ denotes the use of an exceptional $SU(2)$ invariant. 

If one considers additional MIPFs the value of $\Delta$ can only decrease. In this paper
we extend the set of MIPFs of \cite{Schellekens:1989wx} in two ways: by allowing breaking
of world-sheet supersymmetry in the left sector, and by allowing breaking of $SO(10)$ 
to $SU(3)\times SU(2)\times U(1) \times U(1)$. We do indeed observe that  $\Delta$ does go
down in many cases. We have found the following values for $\Delta$:
$$  12, 6, 2, 0$$
plus the aforementioned $\Delta=3$ case, which remains unchanged, and will be discussed in more 
detail below. The largest number of families we have found is 480, and the next-to largest is 360. The distribution
has a huge peak at zero families with over 550.000 distinct MIPFs; there are about 120.000 with 6 and 100.000
with 12 families and 30.000 with 24. Values not divisible by 6 are less common: 25.000 MIPFs with 2, 20.000 with 4 and only 
about 1200 with 3. All values quoted here count mirror pairs as two distinct spectra. Values not divisible by six come
from different tensor combinations than those divisible by 6, so one should not attach too much importance to this
comparison. However, it certainly remains true that the value 3 is disfavored.

These values are found for the exceptional and standard Gepner models as well as for free fermions (where
the only values we found were 0,6,12,18,24 and 48, the same as in \cite{Kiritsis:2008mu}).
The distribution of the values of $\Delta$ over the various CFTs we considered are shown in table \ref{DeltaDist}.




\renewcommand{\arraystretch}{1.1}

\begin{table}[b]
\begin{center}
\vskip .7truecm
\begin{tabular}{|c||c|c|c|}
\hline
\hline
$\Delta$ & Gepner & Gepner (exc.) &  Free Fermions \\
\hline
12 & 10 & 5 &  5   \\
6 & 132 & 33 &  18   \\
3 & 0 & 1 &  0    \\
2 & 5  & 0  & 0     \\
0 & 21  & 20   &  39    \\
\hline
\end{tabular}
\vskip .7truecm
\caption{The number of cases with a given family quantum $\Delta$.}
\label{DeltaDist}
\end{center}
\end{table}

Remarkably, the CFTs that give rise to $\Delta=2$ have total numbers of families
that are {\it never} a multiple of three. For example, the combination $(3,3,3,3,3)$ yields
numbers of values covering all even values up to 50 (with a small, irregular tail reaching
the well-known value 100 for the quintic Calabi-Yau manifold), but with 0, 6, 12, 18, 24, 30, 36, 42 and 48
all missing. 
Furthermore, there is an easily recognizable pattern for the tensor products of the six combinations in this class:
\begin{eqnarray*}
 (6,6,6,6)\\
 (3,3,3,3,3)\\
 (3,6,6,18)\\
 (3,3,18,18)\\
 (3,3,12,33)\\
 (3,3,9,108)\\
\end{eqnarray*}
Obviously, all values of $k$ are divisible by three, and furthermore these are the only combinations
for which that is true.  This is at this moment just an empirical observation, for which an explanation
is still lacking. It is puzzling that it is the value of $k$ rather than $k+2$ that seems to determine the
family quantization properties. The value $k+2$ is a divisor of the simple current order of the minimal model, and 
played an important r\^ole in the discussion of charge quantization. But it is not clear to us what fundamental
property of the CFT the value of $k$ relates to.

\subsubsection{Kazama-Suzuki models}

To see if there is any chance that the foregoing observation has some validity beyond Gepner models and free fermions
we have
examined some old results on Kazama-Suzuki models \cite{Kazama:1988qp}. For these models the unaltered (2,2) spectrum was computed
more than twenty years ago \cite{Font:1989qc,Schellekens:1991sb}. In principle one could treat these models in the same way as we treated the Gepner models: allow
breaking of  world-sheet supersymmetry to (0,2), and allow breaking of $SO(10)$. However, to carry this out we need to know the
exact CFT spectrum of these models, and not just the spectrum modulo integers obtained from the coset construction. Computing
this requires a straightforward, but tedious character decomposition in some cases, and a more tedious and not even
straightforward fixed point resolution procedure in the other cases. To our knowledge, these results are not available
in the literature. To compute the spectra for unmodified (2,2) models just requires the Ramond ground states. Even in that
case, resolving the fixed points turns out to be a difficult matter. Cases without fixed points were considered in \cite{Font:1989qc}, and
with fixed points in \cite{Schellekens:1991sb}. Both papers also studied mixed combinations of Kazama-Suzuki and $N=2$, level $k$
minimal models (which are special examples of Kazama-Suzuki models). The following two observations hold for all
examples listed in these two papers
\begin{itemize}
\item{If minimal models are present with $k$ not divisible by 3, then the number of families {\it is} divisible by 3.}
\item{If minimal models are present and the number of families is {\it not} divisible by 3, then the values of $k$ are divisible
by 3 for {\it all} minimal models in the tensor product.}
\end{itemize}

The property ``$k$ is divisible by 3" seems to extend to Kazama-Suzuki models even if it is not clear what the analog of $k$ is.
For example, the Kazama-Suzuki models\footnote{See  \cite{Font:1989qc} or \cite{Schellekens:1991sb} for the notation.}
$A(2,3,3)$, $A(1,3,3)$, $A(2,2,6)$, $A(1,3,4)$, $A(1,3,9)$, $A(1,3,10)$, $A(1,3,12)$,   $B(3,3)$, $B(3,7)$, $B(3,8)$, $B(3,9)$, $C(3,1)$ and
$D(6,1)$ appear to have that property. All of them give a number of families that is not divisible by three in
combination with minimal models with all $k$ divisible by three. There are a few non-trivial tests of this statement because there are
combinations of these building blocks with each other. Indeed, $B(3,3) \times B(3,3)$, $B(3,3) \times A(1,3,4)$ and 
$A(1,3,4) \times A(1,3,4)$ all yield a number of families not divisible by three. There are a few Kazama-Suzuki models that yield
a number of families not divisible by three all by themselves:
$A(2,2,12)$, $D(7,2)$, $D(9,1)$ and $C(3,4)$. 

One can also identify some Kazama-Suzuki models whose presence implies that the number of families {\it is} divisible by three,
which are therefore analogous to $N=2$ minimal models with $k$ {\it not} divisible by three. These can be identified if they
appear by themselves, or are tensored with building blocks that all belong to the ``$k$ divisible by 3" class. 
They include $A(1,6,7)$, $A(1,5,9)$,  $A(1,4,15)$, $A(1,4,6)$, $A(1,3,6)$, $A(1,3,16)$, $A(1,2,3)$, $A(1,2,2)$, $A(1,2,9)$, $B(3,12)$,
$A(2,3,5)$, $A(3,3,3)$ and  $B(6,6)$. This list provides useful counterexamples to some guesses, for example that
all building blocks in the $A(1,3,k)$ or $B(3,k)$  series belong to the ``$k$ divisible by 3" class.

In Kazama-Suzuki models, if the number of families is divisible by 3, it is almost always also divisible by 6. Three exceptions
were found in  \cite{Font:1989qc}, with 63, 63 and 93 families.

One is led to conjecture that all $N=2$ building blocks come
in two classes: those whose presence in the tensor product {\it enforces} a factor of three in the number of families, and those
whose presence {\it inhibits} a factor of three. Furthermore the enforcers are dominant: the inhibitors can win only
if no enforcers are present. In addition to the Gepner models with $k$ not divisible by three and the Kazama-Suzuki models
listed above, apparently also the free fermionic building blocks are in the enforcer class. 
It is fascinating to see the special r\^ole of the number three in this observation. Twenty years ago one might have
been tempted to link this to the observed family number, despite the additional factor of two one gets in  nearly all cases.
However such a relation is far less convincing in the full string theory landscape. In fact, even in the heterotic
landscape, if we move further away from (2,2) models, we have found\footnote{See \cite{GatoRivera:2009yt} for 
some examples using ``heterotic weight lifting". In a forthcoming paper \cite{ASMHWL} we will present a full analysis of models with heterotic weight lifting,
with family distributions in which all families appear.} that the number three is neither favored nor disfavored. 

It is an interesting question what the origin is of the ubiquitous factor of six (or three) in these constructions. It is clearly {\it not}
the number of colors, since this factor is seen already before considering the Standard Model subgroup of $SO(10)$. A more likely
origin is the number of compactified dimensions. This is indeed true in certain orbifold models, but is not clear why that would
extent (with the exceptions mentioned above) to interacting CFTs.

\subsection{Distributions}

Apart from the two main issues we focus on in this paper, the spectra we have collected
can of course also be used to study other quantities of interest, in particular
how certain non-chiral quantities are distributed. Unlike charge quantization and the number of families, we
do not expect that such quantities can be studied entirely within the context of RCFT, but still a few interesting 
observations can be made.

\begin{figure}[P]
\begin{center}
\includegraphics[width=17cm]{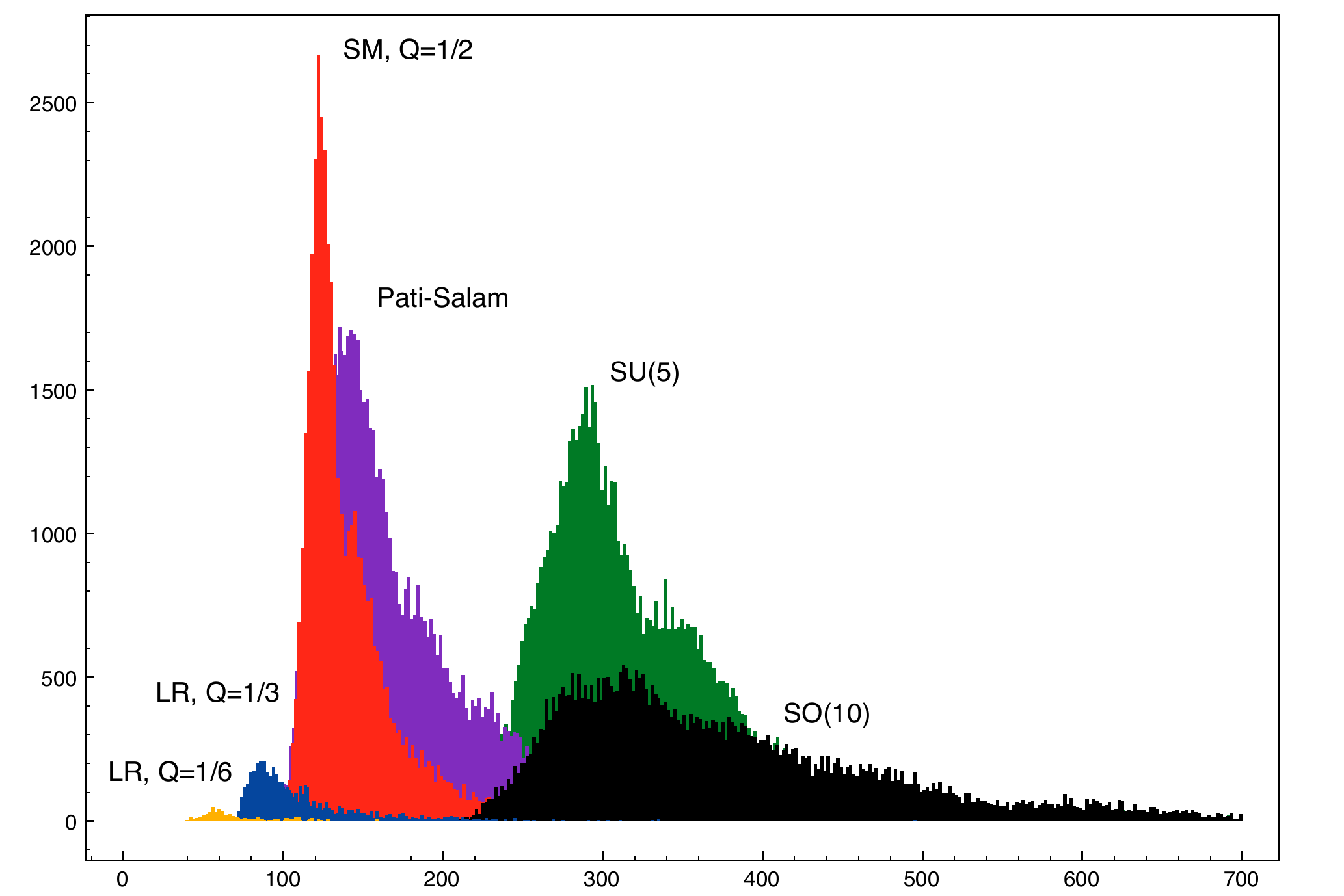}
\caption{Distribution of singlets for Gepner models}
\label{SingGep}
\end{center}
\end{figure}

\begin{figure}[P]
\begin{center}
\includegraphics[width=17cm]{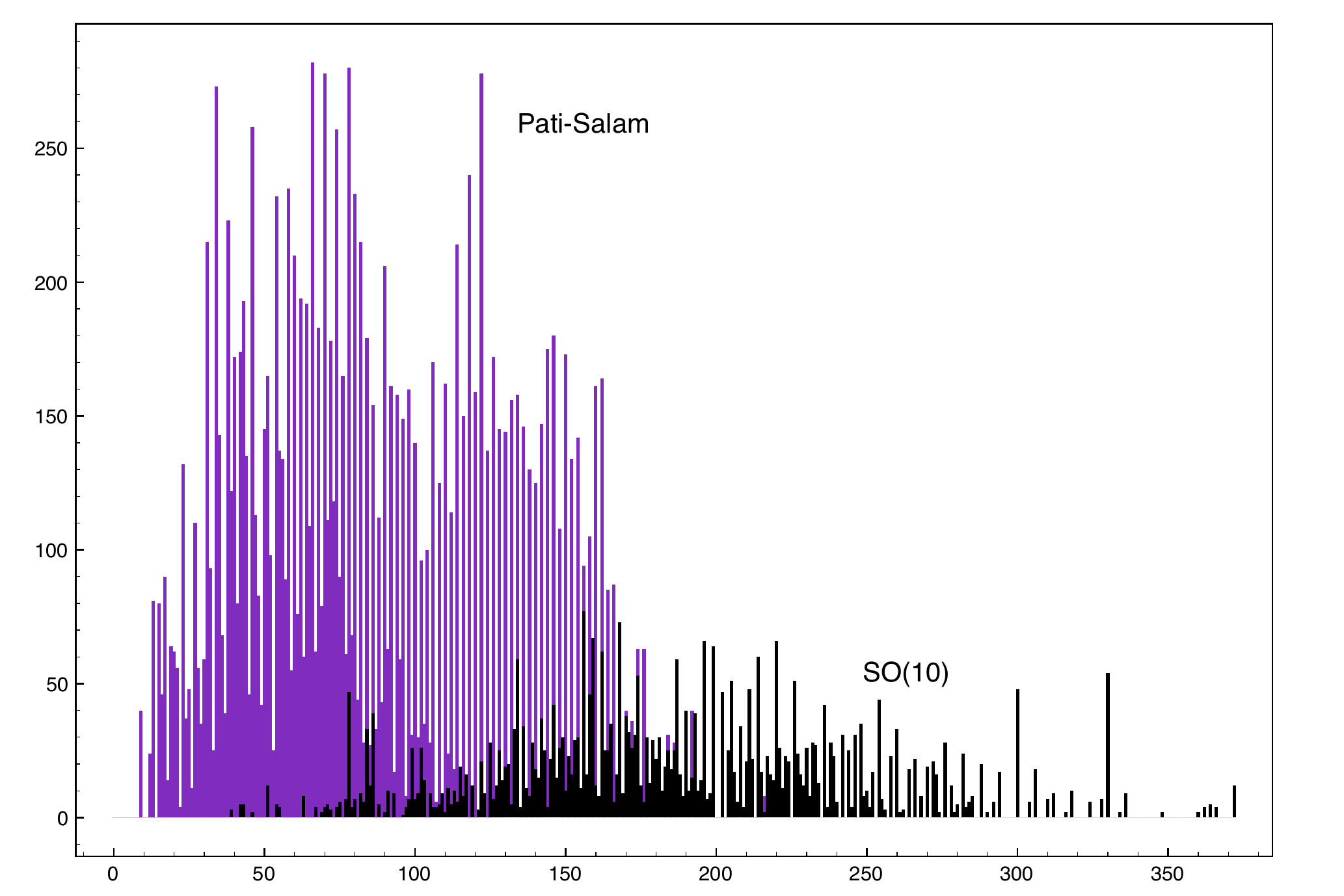}
\caption{Distribution of singlets for Free Fermionic models}
\label{SingFerm}
\end{center}
\end{figure}

\begin{figure}[P]
\begin{center}
\includegraphics[width=12cm]{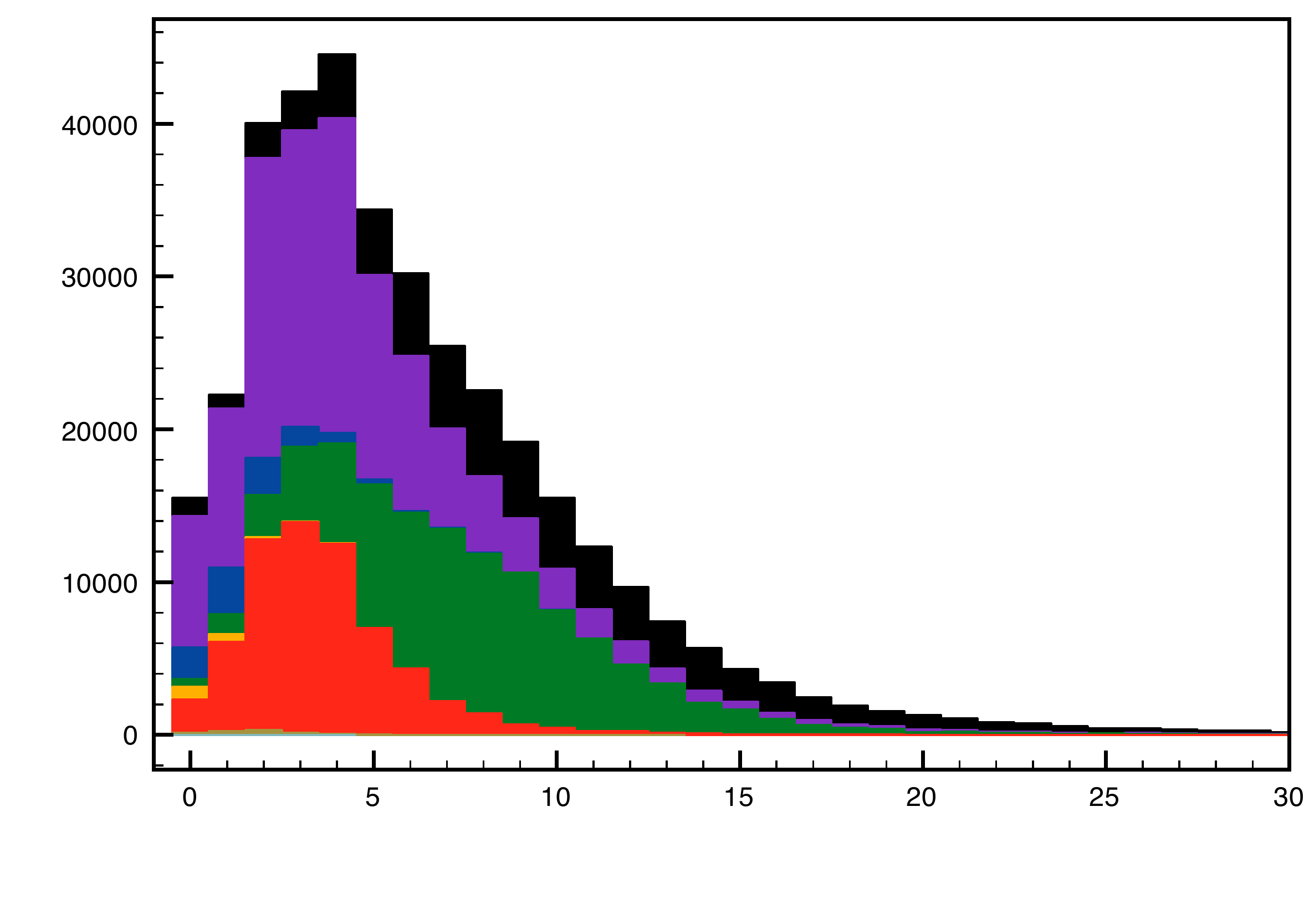}
\caption{Distribution of Q-type mirrors for Gepner models. 
Different types are indicated by different colors stacked on top of each other, ranging from $SO(10)$ on top
to SM, $Q=\frac16$ at the bottom.}
\label{MirrorsQ}
\end{center}
\end{figure}

\begin{figure}[P]
\begin{center}
\includegraphics[width=13cm]{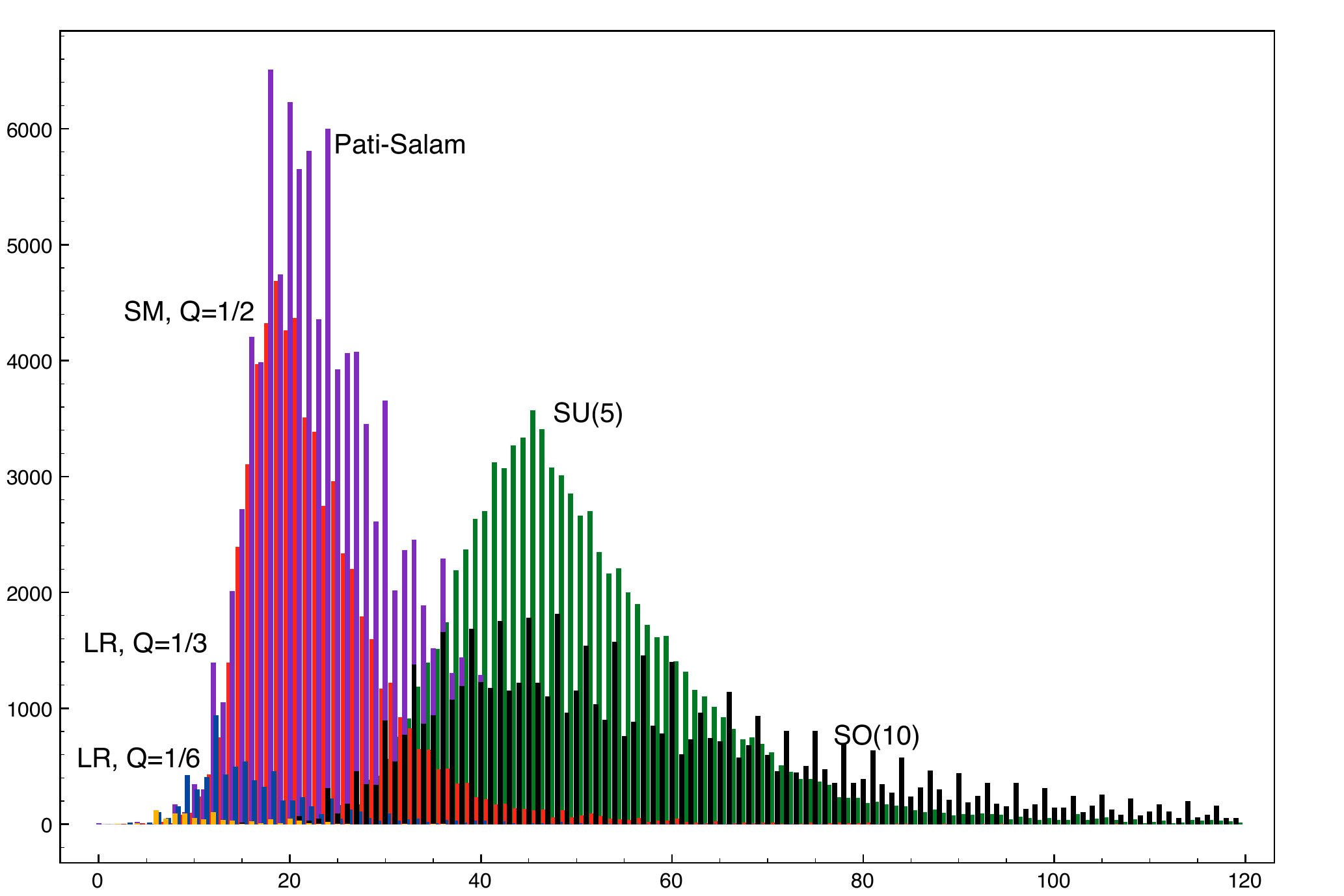}
\caption{Distribution of L-type mirrors for Gepner models.}
\label{MirrorsL}
\end{center}
\end{figure}

\begin{figure}[P]
\begin{center}
\includegraphics[width=13cm]{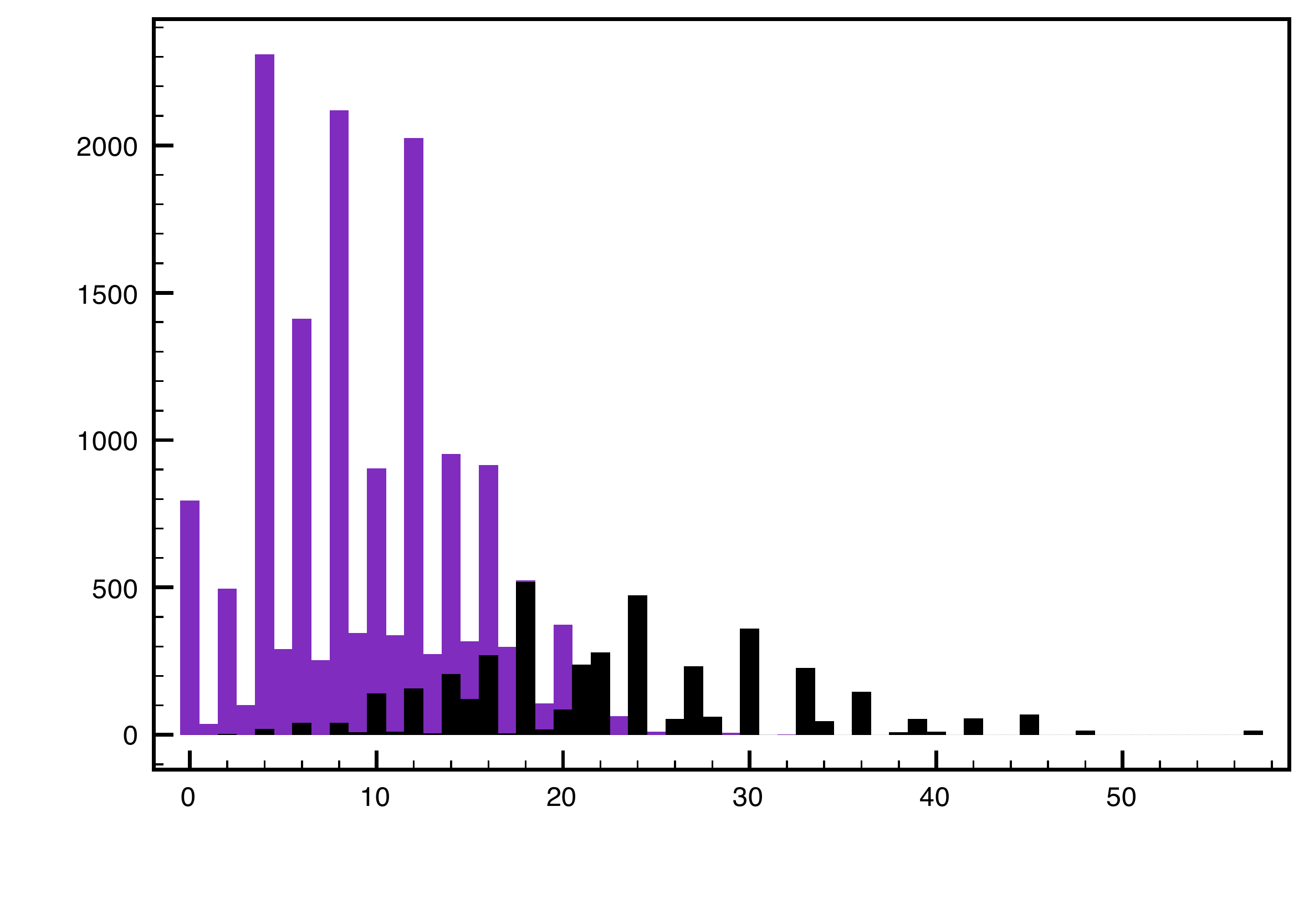}
\caption{Distribution of L-type mirrors for Free Fermionic models.}
\label{LMirrFF}
\end{center}
\end{figure}

In figs.  \ref{SingGep}-\ref{LMirrFF} we plot the distribution of the number of singlets,
of fractional charges, of quark doublet (Q) mirror pairs, and of lepton doublet (L) mirror pairs. In each case
Gepner models and free fermions are compared.
On the vertical axis the number of distinct spectra is indicated.  Note that we are plotting the number of distinct spectra and not their
occurrence frequency. This explains why the SM, Q=$\frac12$ model has a much larger peak than table \ref{Fracs} would suggest.
The distributions are shown for each of the eight types listed in table \ref{TableG}, although some are
too small to be visible. The plot of Q-mirrors is different from the others. For better visibility it has stacked histogram bars (the bars are on top
of each other, not behind each other), so that
each vertical bar is divided into the eight separate contributions, distinguished by colors. In these plots, type 7 ($SO(10)$) appears
at the top of each bar, and type 0 (barely visible) at the bottom.

The distributions of the singlet mirror pairs of types
U and E are nearly indistinguishable from type Q, where D is nearly identical to L, so there is no need to show these. The fact that
the Q, U, E  resp. D, L mirror distributions are almost the same undoubtedly finds its origin in the underlying SO(10) structure. Indeed,
in $SU(5)$ and $SO(10)$ these distributions are automatically identical. 
Furthermore the current $(3,1,10,0)$ present in the model SM, Q=1/2 as well as the Pati-Salam model maps U to E.  
In all other cases the distributions are {\it a priori} not related, but nevertheless we find that  they remain nearly identical, 
with the Q, U, E distributions peaking around or below 5 mirror pairs, and the D, L distributions around 20.

A rather remarkable property of all these distributions, with the exception of the fractional charges, is that they tend towards the phenomenologically correct  value,
zero, as the gauge group is broken to approach the Standard Model. For the Q, U and E mirrors, zero is easily reachable within the
context of RCFT, for D and L mirrors this is clearly much harder, and for singlets this seems impossible. The results for free fermions
show similar features, but they appear to be more affected by RCFT artifacts, in agreement with the expectation that Gepner models
provide a richer and more accurate scan of the heterotic landscape.

\begin{figure}[P]
\begin{center}
\includegraphics[width=13cm]{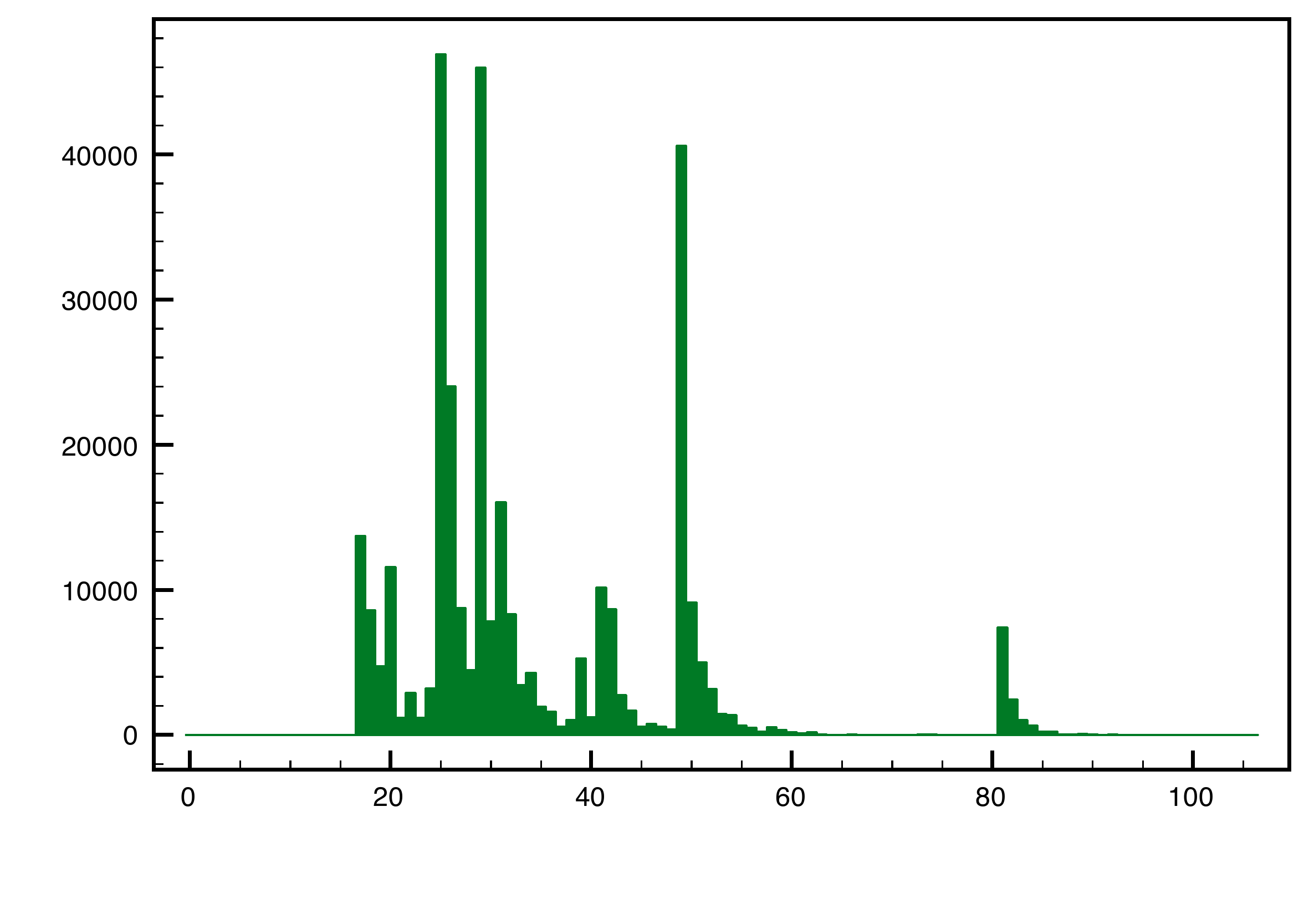}
\caption{Distribution of gauge group dimensions for Gepner models.}
\label{GGep}
\end{center}
\end{figure}

\begin{figure}[P]
\begin{center}
\includegraphics[width=13cm]{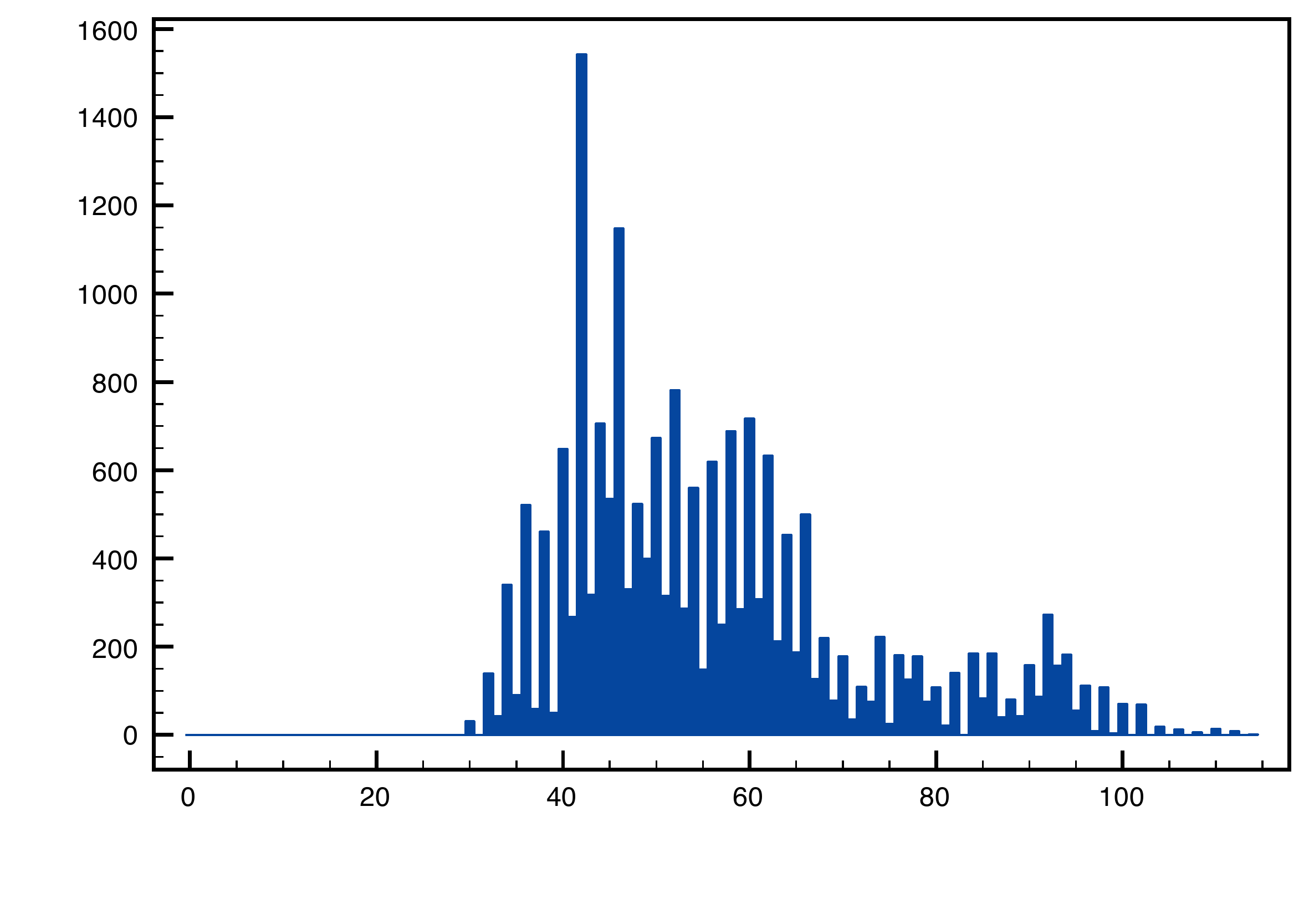}
\caption{Distribution of gauge group dimensions for Free Fermionic models.}
\label{GFerm}
\end{center}
\end{figure}

In figures \ref{GGep} and \ref{GFerm} we show the distribution of the dimension of the gauge group (not including the ``hidden" $E_8$) for 
all distinct spectra with a non-zero number of families. In Gepner models the minimal dimension of the gauge group is
determined by the subgroup of $SO(10)$ that is realized, plus the number of factors in the internal tensor product, each of which
yields at least a $U(1)$. Furthermore the $SO(10)$ (sub)group may be extended into the internal sector, and there may be
factors that live entirely in the internal sector. The best-known $SO(10)$ (sub)group extensions are $E_6$ and $SU(3)^3$. Each of these
absorbs one linear combination of the minimal model $U(1)$'s into the Cartan sub-algebra. We then get the following minimal
dimensions for the various gauge groups if there are $K$ minimal model factors, with $K=4,\ldots 9$.  The last column gives the
absolute minimum for the dimension, which occurs for $K=4$, the most common value.
\begin{eqnarray*}
\hbox{SM}  ~~~~~ &~~~~~   13+K  ~~~~~ &~~~~~  17\\
\hbox{LR}  ~~~~~ &~~~~~    15+K ~~~~~ &~~~~~  19 \\
\hbox{Pati-Salam}  ~~~~~ &~~~~~  21+K  ~~~~~ &~~~~~  25 \\
SU(3)^3  ~~~~~ &~~~~~   23+K ~~~~~ &~~~~~  27\\
SU(5)  ~~~~~ &~~~~~  25+K ~~~~~ &~~~~~ 29  \\
SO(10) ~~~~~ &~~~~~  45+K  ~~~~~ &~~~~~ 49 \\
E_6 ~~~~~ &~~~~~  77+K ~~~~~ &~~~~~  81 \\
\end{eqnarray*}
In figure \ref{GGep}, the $E_6$ and $SO(10)$ peaks are clearly visible, and have a small tail towards larger dimensions due to the
dependence on $K$. Also the $SU(5)$ and Pati-Salam peaks are clearly distinguishable, as
well as the peak at the minimal value for the number of dimensions, 17. The LR peak and the trinification peak are less easily
identifiable; the former may be related to the peak at 20, whereas the latter is hidden in the noise. The results for free fermions in figure \ref{GFerm}
are more difficult to interpret, because in that case large gauge groups are generically present.


\begin{figure}[P]
\begin{center}
\includegraphics[width=13cm]{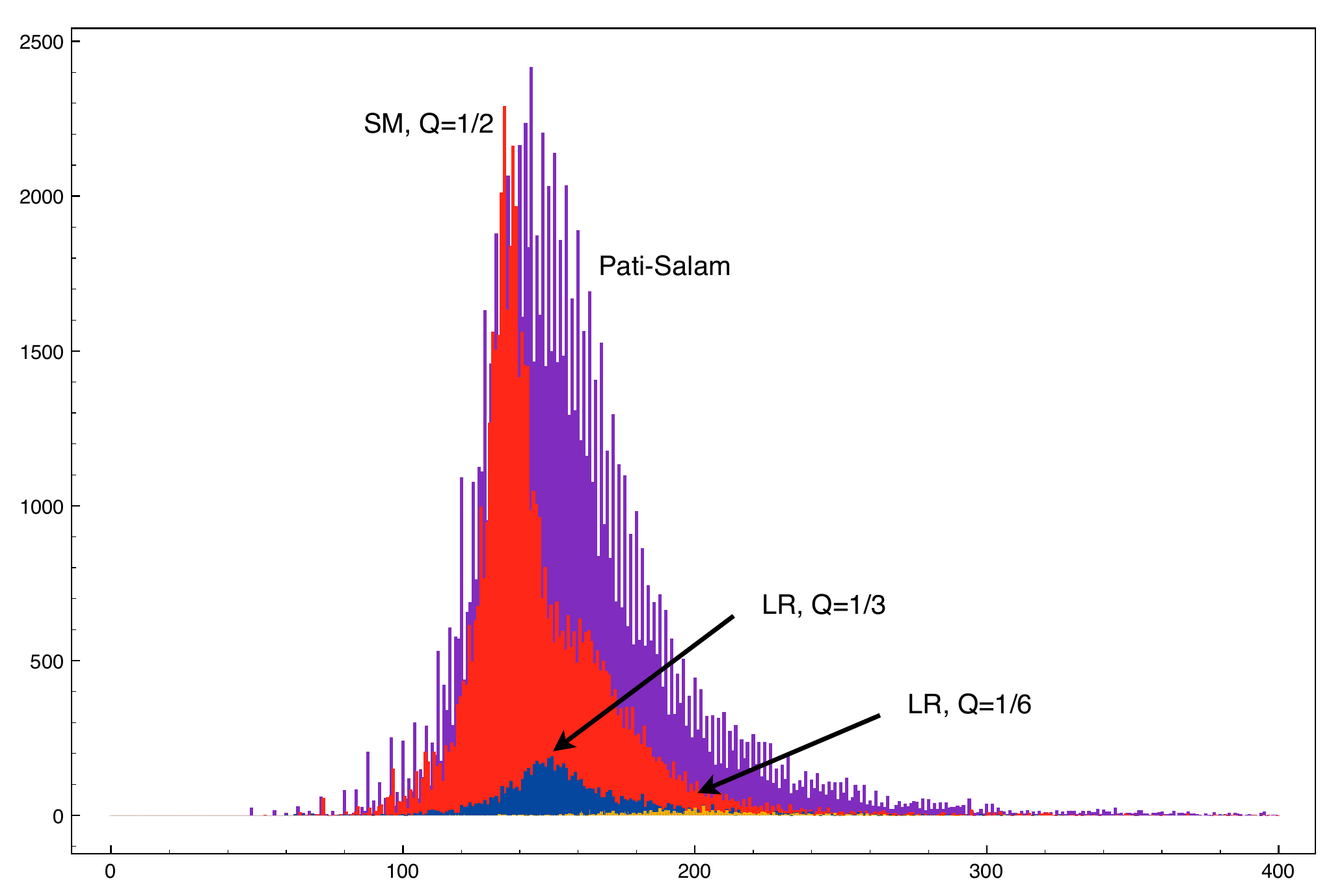}
\caption{Distribution of fractional charges for Gepner models.}
\label{default}
\end{center}
\end{figure}

\begin{figure}[P]
\begin{center}
\includegraphics[width=12cm]{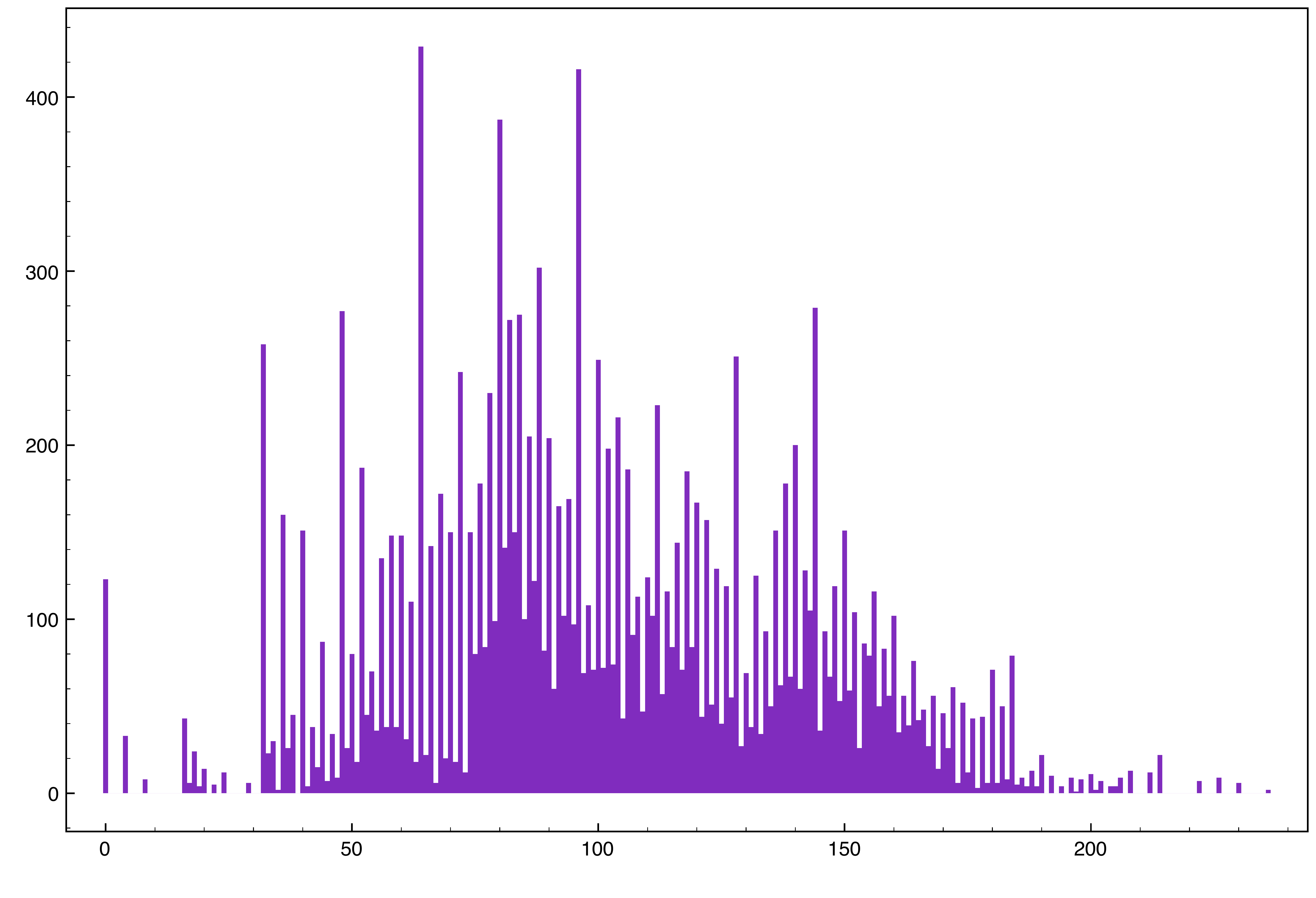}
\caption{Distribution of fractional charges for Free Fermionic models (only Pati-Salam model can occur in this case)}
\label{default}
\end{center}
\end{figure}

\begin{figure}[P]
\begin{center}
\includegraphics[width=12cm]{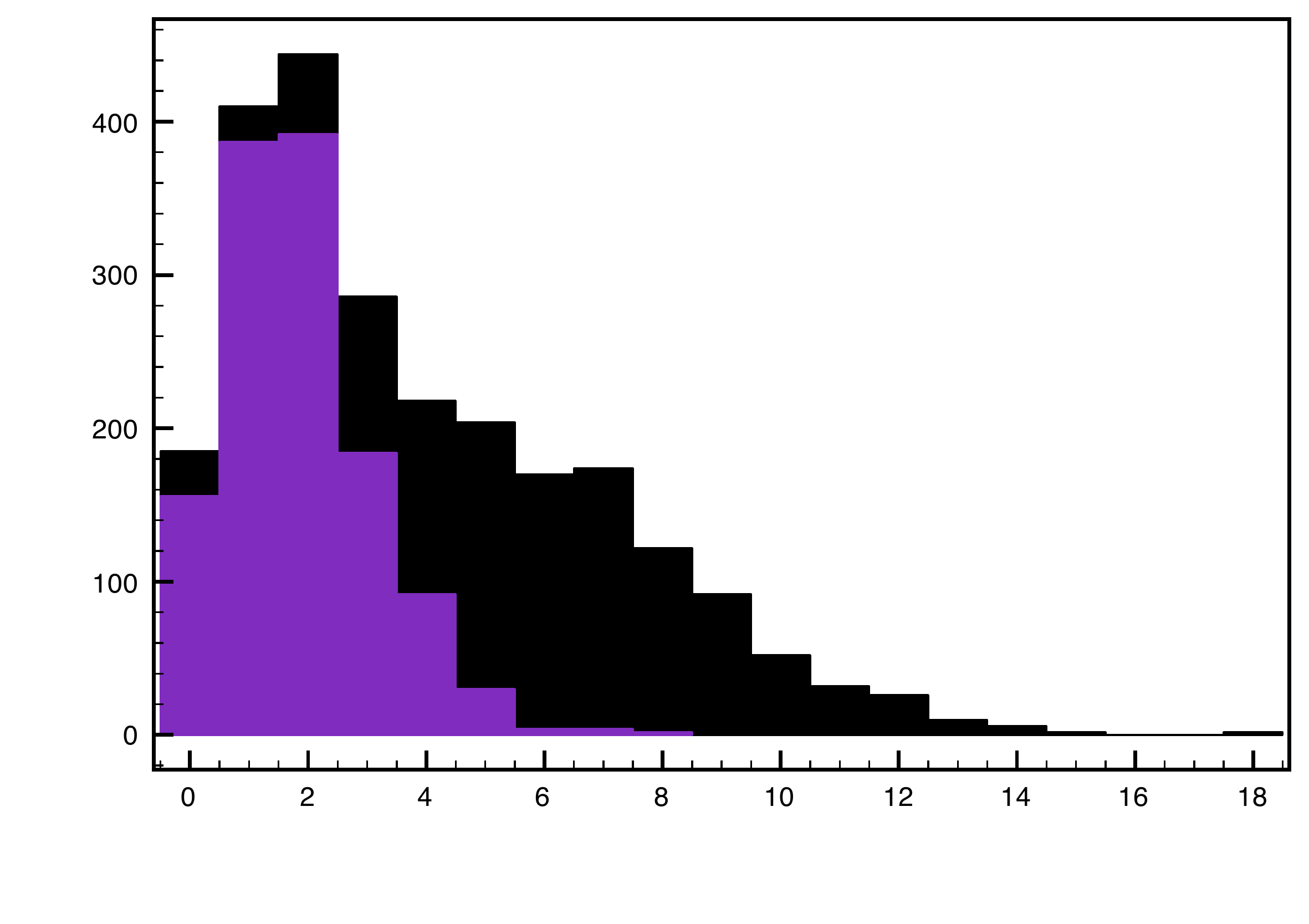}
\caption{Distribution of Q-type mirrors for Free Fermionic models. The color scheme is as in fig. \ref{MirrorsQ}, but only $SO(10)$ and
Pati-Salam models occur. }
\label{default}
\end{center}
\end{figure}

\section{The three family case}

Despite the ubiquitous factors of six in the number of families, we have not found a single new tensor
combination where the greatest common denominator of all  family multiplicities
is reduced to three. This leaves us with the same case first considered by Gepner, the $(1,16^*,16^*,16^*)$ tensor product.
In \cite{Schellekens:1989wx} MIPFs of this tensor product were considered with $(2,2)$, $(1,2)$ and also $(0,2)$ world
sheet supersymmetry, but with unbroken $SO(10)$, and MIPFs constructed as products of at most two single current ones.

In \cite{Schellekens:1989wx} 44 distinct spectra with three families were listed, after modding out mirror symmetry.  In the present study
we have found 1220  three-family spectra, forming precisely 610 mirror pairs. The exact mirror symmetry suggest 
that this set is complete, or very nearly so. Also the complete set of spectra with a non-vanishing number of chiral families has nearly exact mirror symmetry.
It contains 1218 mirror pairs plus one unpaired spectrum, so that almost half
of all the total number of models have three families.

The tensor  product $(1,16^*,16^*,16^*)$ is one of just two where only the gauge groups $LR, Q=1/3$ and $SO(10)$ are possible within
the $c=5$ system ${\cal S}$. These gauge groups may still be extended into the internal sector. In total, there are 58 such extensions,
and the most prominent among them are $E_6$ and $SU(3)^3$ (also called trinification).

\begin{figure}[P]
\begin{center}
\includegraphics[width=12cm]{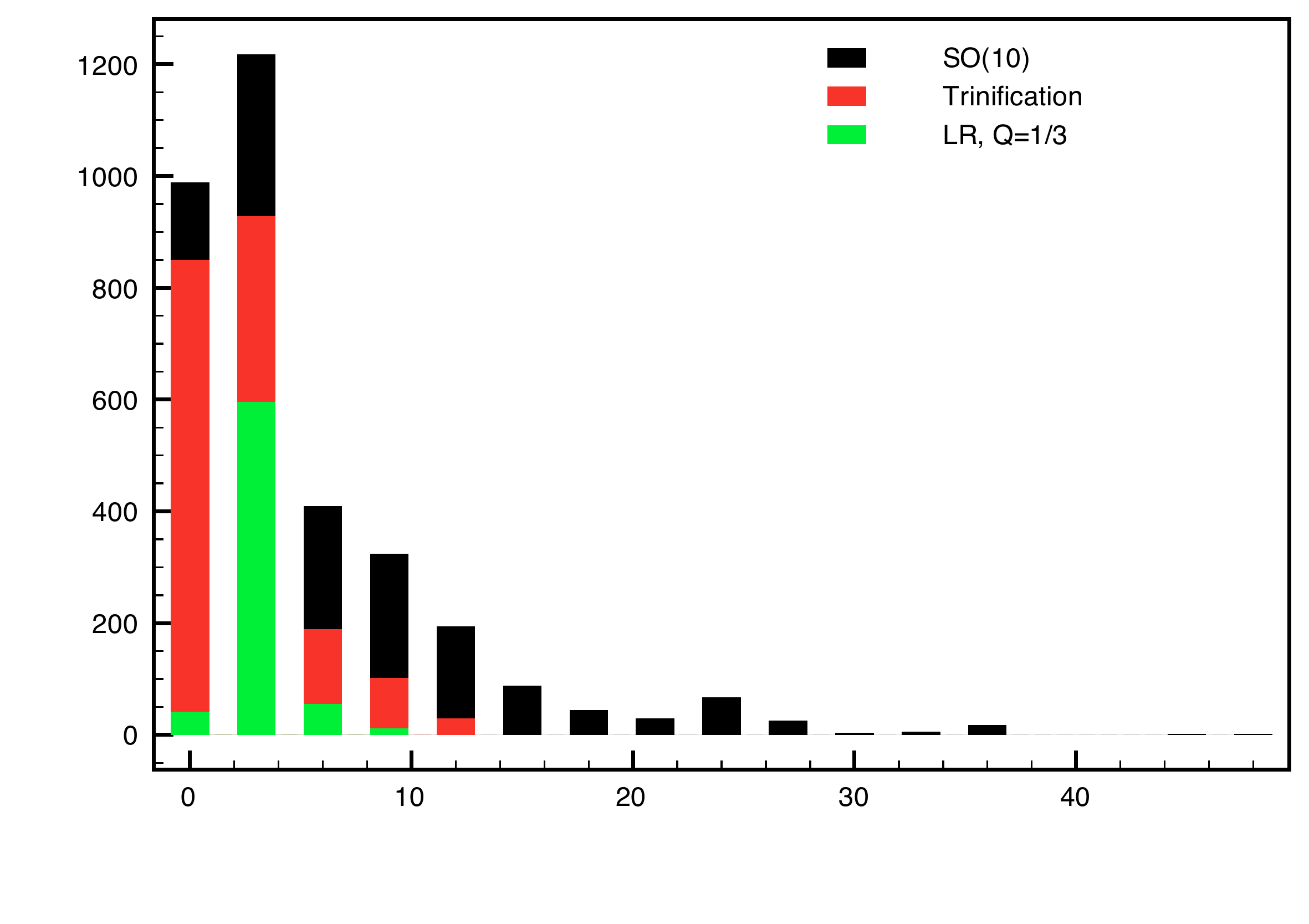}
\caption{Family distribution for the $(1,16^*,16^*,16^*)$ tensor product based on distinct spectra.}
\label{SF}
\end{center}
\end{figure}

\begin{figure}[P]
\begin{center}
\includegraphics[width=12cm]{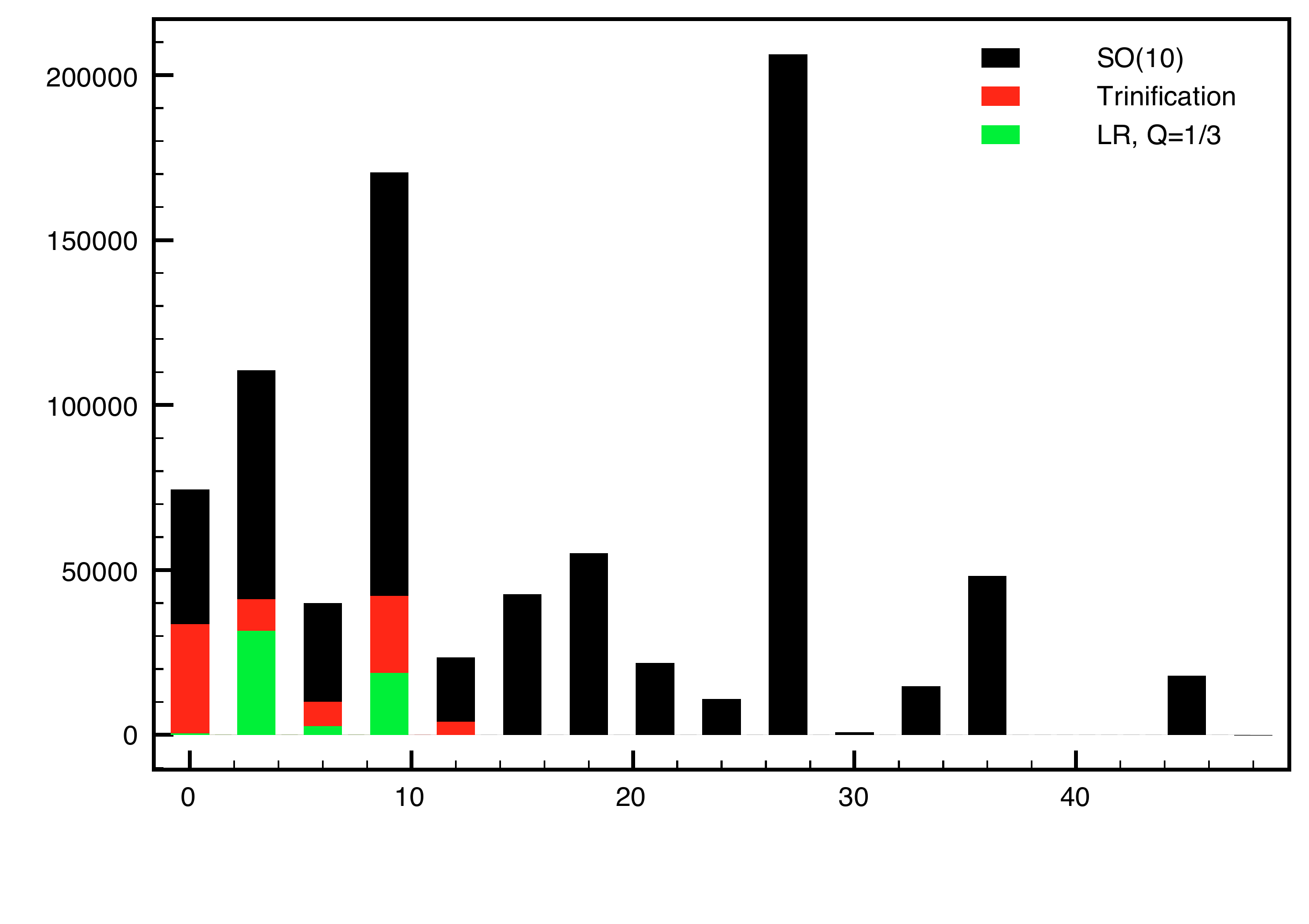}
\caption{Family distribution for the $(1,16^*,16^*,16^*)$ tensor product based on total occurrence frequency.}
\label{STF}
\end{center}
\end{figure}

In fig. \ref{SF} we show the family distribution of all spectra, 
and in fig. \ref{STF} we show the same for the total number of occurrences. Each bar in the histogram is divided into three pieces ({\it i.e} the bars are stacked on top of each other, not hidden behind each other),
indicating, from top to botton, the contribution of $SO(10)$ GUT models (including $E_6$), the contribution of $LR, Q=1/3$ models with
more than 19 gauge bosons, and separately those with exactly 19 gauge bosons. The latter number is the absolute minimum possible
in this tensor combination. It corresponds to the number of generators of $SU(3)\times SU(2)\times SU(2)\times U(1)$ plus 4 $U(1)$'s from
each of the minimal model factors. 
The intermediate case is labelled ``trinification" since 
the majority has 27 gauge bosons, and are trinification models (though not all of them are). The maximal number of families
is 48, and is reached by an $SO(10)$ model that occurs so rarely that its bar is not visible in the plot (the total occurrence
was about 50, compared to about 18000 for the 45 family models). Somewhat atypically, the diagonal MIPF does not have the largest
number of families. It has 27 (8 families and 35 mirror families), and dominates the set with 120000 occurrences, plus 65000 for its mirror.
Trinification models have a maximum of 12 families, whereas for $LR, Q=1/3$ the maximum is 9. It is noteworthy that the number
of families shifts towards smaller numbers as the gauge group approaches the Standard Model. This is presumably in agreement
with expectations, such as those first expressed in \cite{Witten:1985xc}.

A comparison of the two plots illustrates the ``lamppost" effect mentioned in the previous section. It appears evident that the
basic GUT models are overrepresented in occurrence counting. If instead we use spectrum counting (fig. \ref{SF}) we get a 
more or less exponential fall off with the number of families. The slope is considerably smaller than
for orientifold models ({\it c.f.} \cite{Dijkstra:2004cc,Gmeiner:2005vz}). This is a general feature, which we will study in more detail
in a forthcoming paper \cite{ASMHWL}, where family distributions that cover all integers are obtained. It is also remarkable that
the low-family cases are dominated by trinification and $LR, Q=1/3$, and that the three-family case is dominated by spectra
with the smallest possible gauge group. The fact that the family distribution, using spectrum counting, peaks exactly at the
value 3 is interesting, but not typical. In most cases the peak occurs at zero families (except in the class where zero is excluded because
the number of families is not divisible by 3) because there exist algebras that do not have any chiral representations, and hence
only contribute at zero.

 In trinification
models the Standard Model $Y$ generator
and $B-L$ are embedded in $SU(2)_R \times U(1)_{LR}$. The resulting rank-5 algebra is extended to rank 6 by adding a suitable
combination $U(1)_T$ of the $U(1)$ factors from the minimal models. Then $SU(2)_L \times SU(2)_R  \times U(1)_{LR} \times U(1)_T$
is extended to $SU(3)^2$ by extra roots that extend into the internal sector. The Standard Model is embedded in the following
way into the three SU(3) factors
\begin{eqnarray}
(3,1,1)& \rightarrow&  (3,1,0) \\
(1,3,1)& \rightarrow& (1,2,-\frac{1}{6}) + (1,1,\frac13)\\
(1,1,3)& \rightarrow&  2 \times (1,1,\frac13) + (1,1,-\frac{2}{3})\\
\end{eqnarray}
One standard family is contained in the representation $(3,3^*,1) + (3^*,1,3) + (1,3,3^*)$, which in addition to the usual 15 charged
quarks and leptons contains two singlets and a D and L mirror pair. Group theoretically other representations are allowed. In heterotic
strings one may, and in general will, encounter also $(3,1,1), (1,3,1), (1,1,3), (3,3,1), (1,3,3)$ and $(3,1,3)$. All of these give
rise to third-integrally charged particles. 

It is noteworthy that trinification models can be realized
in terms of bi-fundamentals, and therefore have a natural construction in terms of unoriented open strings. They have indeed
been found in RCFT orientifold constructions \cite{adks}, where they constituted less than $1\%$ of the total set of three family models.

\begin{figure}[P]
\begin{center}
\includegraphics[width=13cm]{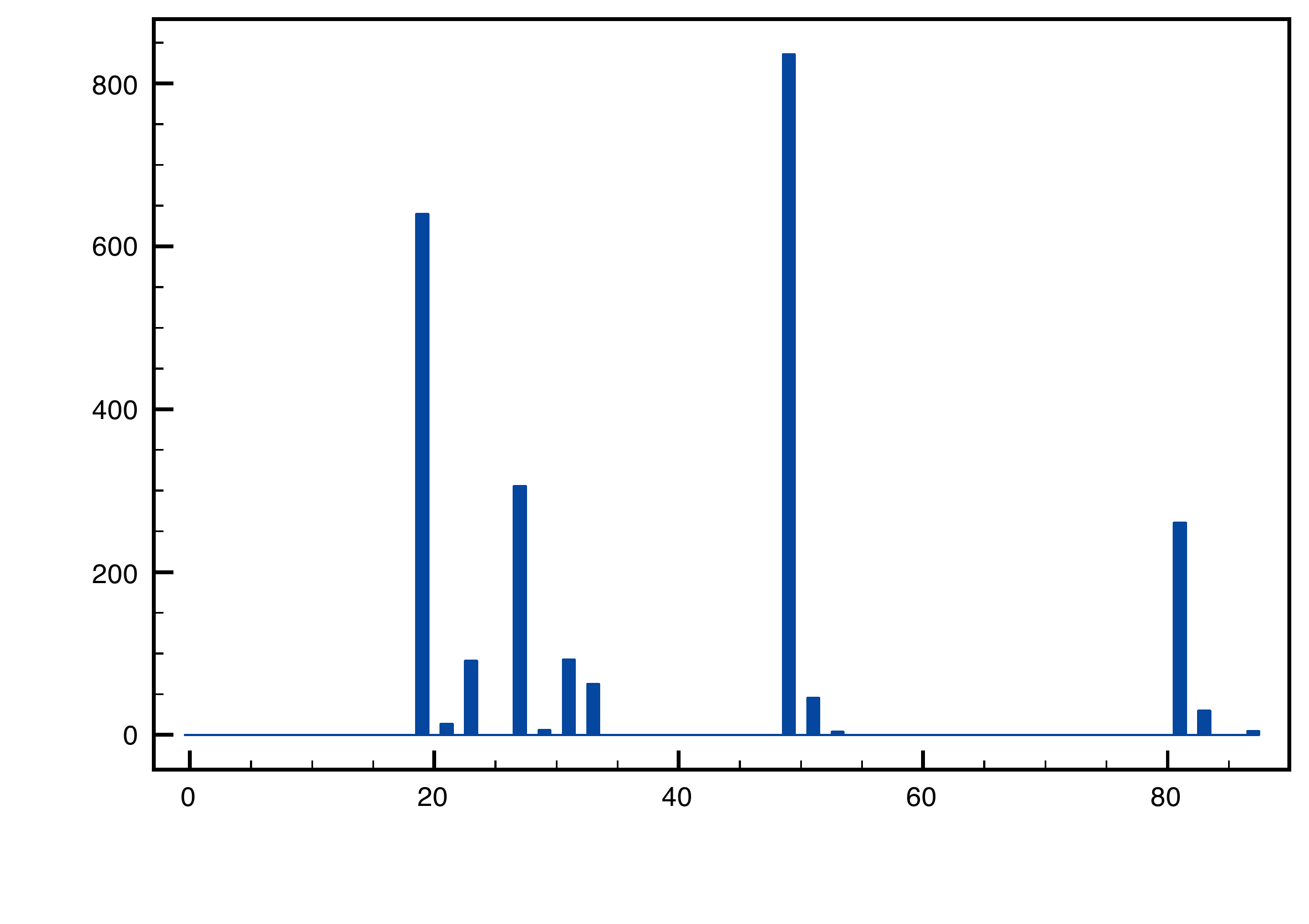}
\caption{Distribution of gauge group dimensions for the $(1,16^*,16^*,16^*)$ tensor product.}
\label{GDE}
\end{center}
\end{figure}

The frequency of trinification models in the complete set is clearly seen in the distribution
of the gauge group dimensions shown in fig. \ref{GDE}. Clearly visible are the peaks for $LR, Q=1/3$ at 19, trinification at 27,
$SO(10)$ at 49 and $E_6$ at 81. 

The distributions of non-chiral states are very similar to those for the full set of Gepner models. The number of fractionally charged
states (the total number of $SU(3)\times SU(2) \times U(1)$ representations, divided by two) peaks at 143, with a minimum at 104. For the $LR, Q=1/3$ models the number of Q,U,E mirrors
peaks at one or two with significant overlap with zero; but the D and L mirrors peak at 9, with a minimum value of 3. For the $SO(10)$
models these distributions peak at larger values, as in the general case, see figs. \ref{MirrorsQ} and \ref{MirrorsL}.

The minimal value for the number of D and L mirrors, 3, occurs precisely for trinification models, where these mirrors are
part of the basic Standard Model family. There are six mirror pairs of spectra with this minimal value, which are very similar: 
They are precisely the six permutations of the following spectrum:
\begin{eqnarray*}
&3 \times [ (3,3^*,1) + (3^*,1,3) + (1,3,3^*)] + 72 \times (1,1,1)\\ + & [2 \times (3,3,1)+ 3 \times (3,1,3) + 3 \times (1,3,3)
+ \hbox{\rm c.c}] \\ + &[19 \times (3,1,1) + 19 \times (1,1,3) + 17 \times (1,3,1)+ \hbox{\rm c.c}]
\end{eqnarray*}
Although this spectrum is free of mirror families (ignoring the mirrors within each family), it does have a large number of
fractionally charged vector-like states, which phenomenologically are much more problematic than mirror fermions. Unfortunately,
we cannot reverse this situation, and allow more mirrors but fewer fractional charges. In all cases the number of fractionally
charged particles is more or less the same as in this example.

The fact that all six permutations of the $SU(3)$ factors occur is intriguing. This permutation symmetry is not directly due to the permutation 
symmetry of the three factors $k=16$. Permutation symmetries among minimal model factors do no affect the spectrum. We did
not investigate if other trinification models in the set we have collected are similarly related. 

Almost half of the three-family models have the theoretical minimal gauge symmetry 
$ SU(3) \times SU(2)_L \times SU(2)_R \times U(1)^5$.  All of these have more mirror fermions than the minimal
trinification model. The minimal number is 12 (as opposed to 6 for the trinification model displayed above), namely one U, six D and five 
L mirrors. 

\section{Conclusions}

We studied interacting CFT constructions of heterotic strings with two main questions in mind: how common are
fractional charges in the spectrum, and how common are three chiral families.  

The answer to the first question is more positive than expected. Mechanisms are available
to limit the fractional charges to half-integer or third-integer. Partly as a result of that, in a substantial fraction
of all cases the fractionally charges are vector-like, and hence may acquire a mass if the spectrum is perturbed. 
Nevertheless, this does not fully restore the original attractiveness of the heterotic string, in our opinion. Absence
of fractional charges is still more beautifully explained in field theoretical GUT models, as well as their string-theoretic
realizations, such as F-theory. But unfortunately those approaches do not offer an explanation why the Standard Model
would emerge via a GUT rather than directly.

The answer to the second
question goes in the right direction, but remains disappointing. Previously, the number of families
was -- with one famous exception -- a multiple of 6 or 4. Now we find that these numbers are typically reduced, but they
are still only multiples of even numbers. Furthermore we find that there are two classes, depending on the tensor product
considered: either all family multiplicities  are a multiple of three, or none of them is divisible by three. In the complete
set of models considered, three families occur still very rarely. However, this problem will be solved in a forthcoming paper \cite{ASMHWL},
where we consider lifted Gepner models, and will arrive at the conclusion that three families occur roughly as frequently as two or four. 

In addition to these questions we studied distributions of various quantities, and found a remarkable tendency of
these distributions in the right direction, in the following sense. As the broken gauge groups gets smaller and hence closer
to $SU(3)\times SU(2) \times U(1)$, all other quantities approach the Standard Model as well. The number of families moves
towards smaller numbers, and a distribution emerges that peaks at zero, and has a (fairly slow) exponential fall-off. With our
present results that conclusion holds for multiples of six or two families, but in \cite{ASMHWL} we will present the same conclusion
for all integers. Furthermore the number of singlets and mirror quarks and leptons peaks at smaller values as the gauge group
approaches the Standard Model. For Q, U and E mirrors the peak is close to zero, and therefore it is not hard to get
exactly zero entirely within RCFT, without giving vevs to scalars. For D and L mirrors zero is harder to reach, but since an
L mirror pair has the quantum numbers of a susy Higgs pair, one might even see that as a positive feature.

\vskip 2.truecm
\noindent
{\bf Acknowledgements:}
\vskip .2in
\noindent
One of us (A.N.S) wishes to thank Nikolas Akerblom for bringing ref. \cite{Schellekens:1987zy} to his attention.  This work has been partially 
supported by funding of the Spanish Ministerio de Ciencia e Innovaci\'on, Research Project
FPA2008-02968, and by the Project CONSOLIDER-INGENIO 2010, Programme CPAN
(CSD2007-00042). The work of A.N.S. has been performed as part of the program
FP 57 of Dutch Foundation for Fundamental Research of Matter (FOM). 

\bibliography{REFS}
\bibliographystyle{lennaert}

\end{document}